\documentclass[twocolumn,prl,superscriptaddress,floats,nobibnotes,notitlepage,citeautoscript]{revtex4-2}
\usepackage{graphicx}
\usepackage[colorlinks=true,urlcolor=blue,linkcolor=blue,citecolor=blue,bookmarks=false]{hyperref}% add hypertext capabilities
\usepackage{epstopdf}
\usepackage[utf8]{inputenc}
\usepackage{amsmath,bm}
\usepackage{bbold}
\usepackage{color,soul}
\usepackage{xstring}
\usepackage{xspace}
\usepackage{amssymb}
\usepackage{time}
\usepackage{bbold}
\usepackage{subfigure}
\usepackage{multirow}
\usepackage{xspace}
\usepackage{url}
\usepackage{amsmath}
\usepackage{amssymb}
\usepackage{ulem}
\usepackage{dsfont}
\usepackage[dvipsnames]{xcolor}
\usepackage{array}
\usepackage{braket}

\def\<{\langle}
\def\>{\rangle}
\DeclareMathOperator{\Tr}{Tr}

\newcommand{\ie}[0]{i.e.\@\xspace}
\newcommand{\eg}[0]{e.g.\@\xspace}

\newcounter{myequation}
\makeatletter
\@addtoreset{equation}{myequation}
\makeatother

\newcounter{myfigure}
\makeatletter
\@addtoreset{figure}{myfigure}
\makeatother

\newcounter{mytable}
\makeatletter
\@addtoreset{table}{mytable}
\makeatother

\newcommand{\im}[0]{\mathrm{i}}

\newcommand{\spin}[1]{\mathbf{S}_{#1}}
\newcommand{\aan}[1]{\mathbf{a}_{#1}^{\phantom\dagger}}
\newcommand{\acr}[1]{\mathbf{a}_{#1}^{\dagger}}
\newcommand{\aantext}[1]{\mathbf{a}_{#1}}
\newcommand{\spinsplit}[1]{\mathbf{s}_{#1}}

\newcommand{\aanc}[1]{{a}_{#1}^{\phantom\dagger}}
\newcommand{\acrc}[1]{{a}_{#1}^{\dagger}}

\newcommand{\banc}[1]{{b}_{#1}^{\phantom\dagger}}
\newcommand{\bcrc}[1]{{b}_{#1}^{\dagger}}
\newcommand{\banctext}[1]{{b}_{#1}}

\newcommand{\vrho}[1]{{\varrho}_{#1}^{\phantom\dagger}}
\newcommand{\vrhodag}[1]{{\varrho}_{#1}^{\dagger}}

\newcommand{\Jang}[1]{\mathbf{J}_{#1}}
\newcommand{\Lang}[1]{\mathbf{L}_{#1}}
\newcommand{\Qbath}[1]{\mathbf{Q}_{#1}}
\newcommand{\Pbath}[1]{\mathbf{P}_{#1}}

\newcommand{\spincoh}[0]{\mathbf{n}}

\newcommand{\Hs}[0]{H_\mathrm{s}}
\newcommand{\Hsb}[0]{H_\mathrm{sb}}
\newcommand{\Hfullint}[0]{\mathcal{H}}
\newcommand{\Hsint}[0]{\mathcal{H}_\mathrm{s}}
\newcommand{\Hret}[0]{\mathcal{H}_\mathrm{ret}}

\newcommand{\Zfull}[0]{Z}
\newcommand{\Zb}[0]{Z_\mathrm{b}}
\newcommand{\Ttime}[0]{\mathcal{T}_\tau}
\DeclareMathOperator{\Trs}{Tr_\mathrm{s}}
\DeclareMathOperator{\Trb}{Tr_\mathrm{b}}

\newcommand{\conf}[0]{\mathcal{C}}

\newcommand{\Ss}[0]{\mathcal{S}_\mathrm{s}}
\newcommand{\Sret}[0]{\mathcal{S}_\mathrm{ret}}
\newcommand{\Stop}[0]{\mathcal{S}_\mathrm{top}}

\newcommand{\phivec}[0]{\boldsymbol{\phi}}

\newcommand{\omc}[0]{\omega_\mathrm{c}}

\newcommand{\alphac}[0]{\alpha_\mathrm{c}}

\newcommand{\Rspin}[0]{R_\mathrm{spin}}
\newcommand{\Rstring}[0]{R_\mathrm{string}}
\newcommand{\Cspin}[0]{C_\mathrm{spin}}
\newcommand{\Cstring}[0]{C_\mathrm{string}}

\newcommand{\Dstring}[0]{\Delta_\mathrm{string}}
\newcommand{\De}[0]{\Delta_\mathrm{e}}
\newcommand{\Db}[0]{\Delta_\mathrm{b}}

\newcommand{\dimred}[0]{\mathsf{d}}

\begin{document}

\title{Dissipation-induced bulk and boundary criticality in the Haldane chain}

\author{Zhenjiu Wang}
\email{wangzj@lzu.edu.cn}
\affiliation{Lanzhou Center for Theoretical Physics, Key Laboratory of Quantum Theory and Applications of MoE, Key Laboratory of Theoretical Physics of Gansu Province, and School of Physical Science and Technology, Lanzhou University, Lanzhou, Gansu 730000, China.}

\author{Manuel Weber} 
\email{mweber.physics@gmail.com}
\affiliation{Institut f\"{u}r Theoretische Physik and W\"{u}rzburg-Dresden Cluster of Excellence ctd.qmat, Technische Universit\"{a}t Dresden, 01062 Dresden, Germany}
\affiliation{Laboratoire L\'{e}on Brillouin, CEA, CNRS, Universit\'{e} Paris-Saclay, CEA Saclay, 91191 Gif-sur-Yvette, France}

\begin{abstract}
We study the Haldane chain coupled to a dissipative ohmic bath. Using large-scale quantum Monte Carlo simulations, we identify a second-order quantum phase transition into an antiferromagnetic state with spontaneously broken SO(3) symmetry that is governed by an interacting fixed point with dynamical exponent $z\approx 2$. We derive a generalized string order parameter which indicates that the symmetry-protected topological ground state of the Haldane chain is stable for weak dissipation and develops a nontrivial scaling dimension at criticality. In particular, its topological edge modes display nontrivial boundary criticality that is distinct from an equivalent transition out of a trivial state; for the latter, we perform an accurate $\epsilon$ expansion of a dissipative $\phi^4$ theory with ordinary boundary conditions. Our setup realizes a spin impurity in a critical yet nonconformal antiferromagnet and paves the way for continuously tunable boundary criticality controlled by dissipation.
\end{abstract}
\maketitle

\textit{Introduction.}---%
Dissipation is a powerful mechanism for realizing novel phases of matter and quantum phase transitions in low dimensions.
By suppressing quantum fluctuations, the memory of the environment can fundamentally modify the low-energy physics.
In this way, dissipation can even lift the restrictions of the Mermin-Wagner theorem 
by damping the Goldstone fluctuations 
that otherwise would destroy long-range order associated with continuous symmetry breaking. 
Such unconventional orders appear in a variety of settings---including Josephson-junction arrays \cite{FAZIO2001235, PhysRevLett.56.2303, Sachdev_2004_Nano, PhysRevB.75.014522}, 
open Luttinger liquids \cite{PhysRevLett.79.4629, PhysRevLett.97.076401, PhysRevB.80.214515},
quantum spin chains in metallic environments \cite{Sachdev_2004_NFL, Werner_2005, PhysRevB.86.035455, PhysRevLett.129.056402, danu2022_chain_metal, PhysRevResearch.5.043270, danu2025phasesphasetransitionss32, lu2026deconfinedquantumcriticalpoint},
superfluidity %on comb geometries
in hybrid-dimensional geometries 
\cite{PhysRevB.109.L100502, PhysRevB.110.115145, dljc-j3z7, cazalilla2026superconductormetaltransitiononedimensionalinteracting}, 
and even inspired classical analogs \cite{1jkm-kybr}---and are typically reached 
through nonconformal quantum phase transitions \cite{Sachdev_2004_NFL, Werner_2005}.
However, 
it remains largely unexplored how dissipative quantum criticality is modified when it originates from a symmetry-protected topological (SPT) phase with protected boundary degrees of freedom.

Quantum spin chains provide an ideal platform for investigating this interplay. 
Originally, the effects of dissipation were studied for the $O(N)$ nonlinear sigma model.
Its quantum phase transition from a gapped ground state to a dissipation-induced long-range-ordered phase is well described by a powerful $\epsilon$-expansion technique \cite{Sachdev_2004_NFL, Werner_2005}. 
However, it neglects the topological $\theta$ term which originates from nontrivial spin-Berry phases.
For half-integer spin quantum numbers $S$, $\theta = \pi$ renders the ground state of the isolated spin chain critical \cite{Haldane83, HALDANE1983464} and therefore leads to fundamentally different criticality under dissipation \cite{PhysRevLett.129.056402, PhysRevResearch.5.043270}.
The situation is more subtle for integer $S$:
Although the topological term is trivial within the bulk, the pure nonlinear sigma model
does not capture that the ground state of the $S=1$ Haldane chain is an SPT
state \cite{PhysRevB.80.155131, Pollmann10, PhysRevB.83.035107, PhysRevB.85.075125}
 characterized by a nontrivial string order parameter \cite{PhysRevB.40.4709, PhysRevB.45.304} and fractionalized $S=1/2$ edge modes \cite{PhysRevLett.59.799}. It remains an open question how these SPT features affect the dissipation-induced criticality in the bulk \textit{and} at the boundary.

Magnetic adatoms on a metallic substrate \cite{Toskovic16} offer a promising route to realizing dissipative spin chains. 
To leading order in Hertz-Millis theory \cite{Hertz76, Millis93}, the conduction electrons act as a local ohmic bath. 
While this coupling immediately stabilizes antiferromagnetic (AFM) order in an $S=1/2$ chain \cite{PhysRevLett.129.056402, danu2022_chain_metal}, partial Kondo screening of $S=3/2$ adatoms realizes a dissipative Haldane chain that undergoes the ordering transition only beyond a finite critical coupling \cite{danu2025phasesphasetransitionss32}.
Such a setup gives access to the topological edge states but neither bulk nor boundary criticality have been investigated due to limited system sizes accessible to fermionic Monte Carlo.

Protected edge states can induce novel boundary universality classes through their coupling to critical bulk fluctuations \cite{Grover12, Long_Zhang_2017,Zhu_2021,Zhe_Wang_2023, ml88-hbd5, toldin2025extraordinarytransitionedgecorrelated, PhysRevX.7.041048, PhysRevX.11.041059}. 
Even for conventional O($N$) models, 
the classification of boundary critical phenomena \cite{doi:10.1142/S0217979297001751} 
has only recently been completed \cite{Metlitski_2022, PhysRevLett.128.215701}, 
renewing interest in boundary conformal field theories, \eg, in the context of spin impurities in critical O($N$) antiferromagnets \cite{SachdevImpurity99, Vojta_2000, Cuomo_2022, 2025arXiv250814963K, 2026arXiv260407554S}. 
While the fate of the SPT state under strong dissipation is a subject of active investigation \cite{2019iSci...21..241H, PhysRevLett.126.237201}, 
dissipation-induced boundary criticality has received little attention. Apart from the conductance of nanowires analyzed within the $\epsilon$ expansion \cite{Sachdev_2004_Nano}, a systematic characterization of boundary critical behavior is still lacking, even for dissipative O($N$) models.
Indeed, already for a trivial bulk theory, the absence of Lorentz invariance promises access to novel surface criticality beyond boundary conformal field theory.

In this Letter, we investigate the dissipation-induced bulk and boundary criticality of the Haldane chain. To isolate the role of the SPT phase, we systematically compare our results with a topologically trivial phase. 
Using large-scale quantum Monte Carlo (QMC) simulations, we show that both models undergo a bulk transition into an AFM phase whose critical exponents are well described by the $\epsilon$ expansion \cite{Sachdev_2004_NFL}.
Nevertheless, a generalized string order parameter demonstrates that the SPT phase remains intact up to the critical point, where it acquires a nontrivial scaling dimension.
The boundary critical behavior, however, is fundamentally different.
 While the trivial chain is well described by the $\epsilon$ expansion,
 which we extend to surface criticality, the  edge states of the Haldane chain induce novel boundary exponents. The latter resembles a spin impurity coupled to a critical antiferromagnet, but here the critical bulk is intrinsically nonconformal. In the future, dissipative spin chains offer a unique platform for exploring boundary criticality with a continuously tunable effective bath dimension, enabling access to quantum critical regimes that are difficult to realize in fixed spatial dimensions.

\textit{Model.}---%
We consider an $S=1$ quantum spin chain
\begin{equation}\label{eq:def_dimerized}
   \Hs = J \sum_{   i  }  \left[1 - (-1)^{ i } \delta \right] \spin{i}  \cdot \spin{i + 1} 
\end{equation}  
with an antiferromagnetic exchange coupling $J=1$ and spin-1 operators $\spin{i}$ defined on lattice sites $i\in\{1, \dots, L\}$.
We include an additional dimerization parameter $\delta$ that allows us to distinguish between two gapped phases: 
the nontrivial Haldane phase at weak $\delta$ and a trivial insulator at large $\delta$. Here we choose $\delta=0$ and $\delta=0.4$, respectively.

To investigate the effect of an environment on the two gapped ground states and, in particular, 
how their trivial or nontrivial nature affects the outcome,
we supplement our Hamiltonian $ H = \Hs + H_\mathrm{b} + \Hsb$ with a coupling to a bath of harmonic oscillators,
\begin{equation}
  H_\mathrm{b} + \Hsb = \sum_{ iq } \omega_{q} \, \acr{iq} 
  \cdot \aan{iq} + 
  \sum_{iq} \lambda_{q} \, \big( \acr{iq} + \aan{iq} \big)  \cdot \spin{i}  \, .
\label{Eq:def_Hsb}  
\end{equation}
Each spin $\spin{i}$ is coupled to an independent bath of three-component vector bosons described by annihilation operators $\aantext{iq}$ and a set of modes $q$. This coupling conserves a global SO(3) symmetry of system plus bath. We define the spectral density
$J( \omega ) = \pi \sum_q \lambda^2_q  \, \delta( \omega - \omega_q )$ of the bath which, in the continuum limit, obeys the power-law form
$
    J(\omega)  = 2 \pi \alpha J^{1-s} \omega^s$
for $0 < \omega < \omc$.
Here, $\alpha$ denotes the dissipation strength and $\omc / J =10$ the frequency cutoff.
We consider an ohmic bath with exponent $s=1$.

Using the path integral, the bath can be integrated out exactly, generating an attractive spin interaction \cite{SM2026}
\begin{align}
\label{eq:sret}
\Hret = - \iint_0^\beta  d \tau \, d \tau' \sum_i  K( \tau - \tau' ) \,  \spin{i}(\tau)
\cdot \spin{i}(\tau') 
\end{align}
that is nonlocal in imaginary time and mediated by the bath propagator which obeys $K(\tau) \propto \alpha/\tau^{1+s}$ for $\omc \tau \gg 1$.
Here $\beta=1/T$ is the inverse temperature.

\textit{Low-energy theory.}---%
The isolated quantum spin chain is described by the nonlinear sigma model
$\Ss = \Stop + \frac{1}{2}\int d\tau \int d r \left[ \chi_\perp \, (\partial_\tau \spincoh(r,\tau) )^2 + \rho_\mathrm{s} \, (\partial_r \spincoh(r,\tau))^2\right]$, 
where $\spincoh$ is an $N=3$ component unit vector. 
The additional topological term $\Stop$ does not affect the bulk properties of the two gapped ground states of $\Hs$, but becomes relevant at the boundaries of the Haldane phase \cite{T-K-Ng} where it induces localized edge states with fractional spin quantum number $S/2$. 
In accordance with Eq.~\eqref{eq:sret}, the coupling to an ohmic bath leads to an additional retarded interaction
$\Sret \propto - \alpha \int d\tau \int d\tau' \int dr \, \spincoh(r,\tau) \cdot \spincoh(r,\tau') / |\tau-\tau'|^2$.

The effect of an ohmic bath on the  bulk theory has been examined using an O($N$)-symmetric dissipative $\phi^4$ theory \cite{Sachdev_2004_NFL}
\begin{align}
\nonumber
\mathcal{S}
    = &\frac{1}{2} \int d\omega \int d^d\mathbf{k} \left[\mathbf{k}^2 + \left| \omega \right| + r \right] \left|\phivec(\mathbf{k},\omega)\right|^2
    \\
    &+ \frac{g}{4!} \int d\tau \int d^d\mathbf{r} \left[\phivec(\mathbf{r},\tau)^2\right]^2
\label{eq:Seps}
\end{align}
by performing an $\epsilon$ expansion about the upper critical dimension $d=2$. For any $N$, it predicts a second-order quantum phase transition from a disordered ground state to a long-range-ordered phase
\cite{Sachdev_2004_NFL, footnote_decay}. Although the latter spontaneously breaks the continuous O($N$) symmetry for $N\geq 2$, this is not in violation of the Mermin-Wagner theorem because of the long-range retarded interaction. Bulk critical exponents have been calculated up to $\mathcal{O}(\epsilon^2)$ \cite{Sachdev_2004_NFL, Sachdev_2004_Nano}. For $N=1$ and $N=2$, they have been confirmed to be in good agreement with classical Monte Carlo simulations of the nonlinear sigma model without topological term \cite{Werner_2005_ising, Werner_2005}. For our case of $N=3$, the $\epsilon$ expansion predicts a correlation-length exponent of $\nu=0.688$, an anomalous dimension of $\eta=0.022$, and a dynamical critical exponent of $z=2-\eta=1.978$. We have also extended the $\epsilon$ expansion to calculate the surface critical exponents \cite{SM2026}. All results for $N=3$ are collected in Table~\ref{tab:exponents}.

\begin{table}[t]
\caption{\label{tab:exponents}%
Bulk and boundary critical exponents for the Haldane and dimerized chains compared with the $\epsilon$ expansion. 
}
\begin{ruledtabular}
\begin{tabular}{c|c|c|c}
Exponents &
Haldane chain &
dimerized chain &
$\epsilon$ expansion \\
\colrule
$\nu$ & 0.697(8) & 0.71(2) & 0.688\\
$\eta$ &  0.01(2) & 0.03(2) & 0.022\\
$z+\eta$ & 1.99(2) & 2.02(2) & 2\\
$\Delta_\mathrm{string}$ & 0.154(5) & - & -\\
\colrule
$\eta_\parallel$ & $-0.28(1)$ & 1.27(3) & 1.288
\end{tabular}
\end{ruledtabular}
\end{table}

\textit{QMC method.}---%
For our simulations of the dissipative quantum spin chain, we use an exact QMC method for retarded interactions \cite{Weber17, PhysRevB.105.165129} that takes advantage of the fact that the bosonic bath can be integrated out analytically.
The resulting partition function $\Zfull = \Zb \Trs \Ttime \, e^{-\Hfullint}$
is expanded in the full interaction $\Hfullint = \Hsint + \Hret$,
in close analogy to the stochastic series expansion. For efficient sampling, we
use the global directed-loop updates for $\Hsint$
\cite{Sandvik99b, Syljuasen02}
and the wormhole moves for $\Hret$
\cite{PhysRevB.105.165129}. For details see the Supplemental Material (SM) \cite{SM2026}.

For our analysis, we calculate the spin-spin correlation functions
$
C(i,j;\tau) = \langle S_i^x(\tau) S_j^x(0)  \rangle
$
from which we define
the equal-time correlations $C(i,j) = C(i,j;\tau=0)$, the static susceptibilities $\chi(i,j) = \int d\tau \, C(i,j;\tau)$, and,
for periodic boundary conditions, their Fourier transform
$\chi(q)= \frac{1}{L}\sum_{ij} e^{\im q (i-j)} \chi(i,j)$.

\textit{Bulk criticality.}---%
We first consider the effects of ohmic dissipation on the Haldane phase under periodic boundary conditions.
The $\epsilon$ expansion predicts a bulk quantum phase transition with dynamical critical exponent $z\approx 2$.
To access the critical point in our finite-size scaling analyis, we scale inverse temperatures as $\beta J = L^2/48$ but also confirm $z\approx 2$ independently below.

The quantum phase transition to long-range AFM order
can be detected using the correlation ratio
\begin{equation}
\label{eq:Rspin}
   \Rspin(L)  =  1 - \frac{  \chi( q= \pi + \delta q ) }{  \chi (q=\pi) } \, ,   \qquad \delta q = 2\pi / L \, .
\end{equation} 
Here, $q=\pi$ is the AFM ordering vector and $\delta q$ the momentum resolution for a given system size $L$. 
$\Rspin(L)$ is defined in such a way that it scales to zero in the disordered phase, to one in the ordered phase, and becomes scale invariant at a quantum critical point
\footnote{Note that the use of $\chi(q)$ instead of $C(q)$ significantly reduces finite-size corrections in our crossing analysis.}. 
This is confirmed by
Fig.~\ref{fig:bulkcriticality}(a) which shows $R_\mathrm{spin}(L)$ as a function of the dissipation strength $\alpha$.
\begin{figure}[t]
  \centering
  \includegraphics[width=\linewidth]{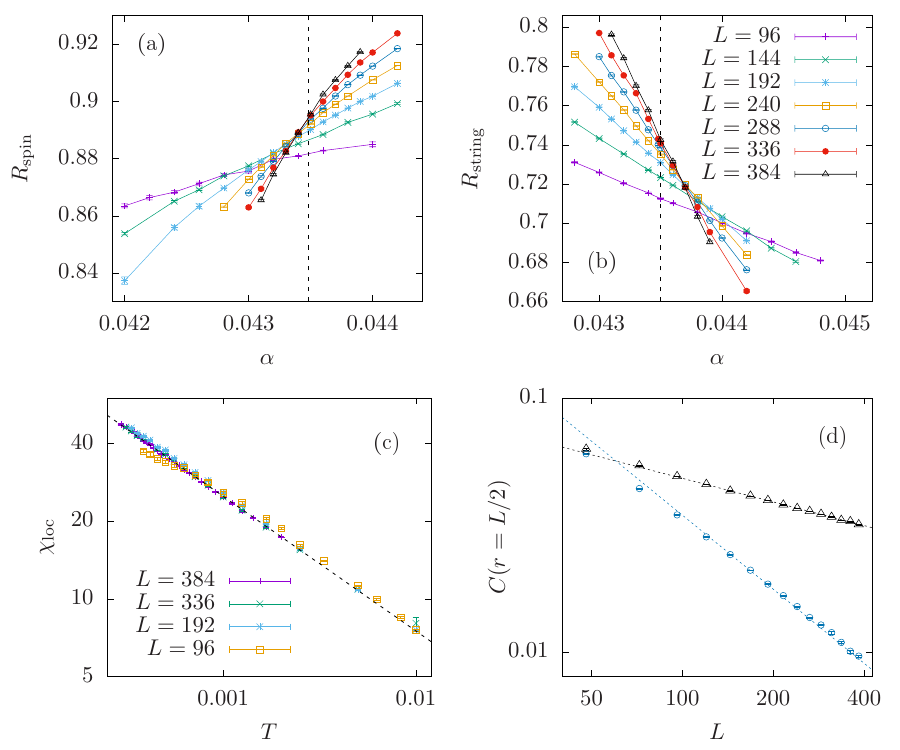}
  \caption{\label{fig:bulkcriticality}%
  Bulk criticality of the Haldane chain. Correlation ratios calculated from the (a) spin and (b) string order parameters as a function of $\alpha$ and for different systems sizes $L$. The dashed lines indicate our extrapolated critical couplings \cite{SM2026}.
 (c) Local susceptibility at $\alpha = \alphac$ as function of temperature for various $L$. (d) Spin and string correlation function at $r=L/2$ as a function of $L$. The dashed lines in (c) and (d) are power law fits to the asymptotic behavior \cite{SM2026}.
  }
  \end{figure}
The crossings of $R_\mathrm{spin}(\alpha)$ for data pairs $(L-\Delta L,L)$ define a pseudocritical coupling that 
can be extrapolated to obtain the critical coupling  $\alpha_c=0.04348(2)$ between the disordered and the AFM-ordered phase \cite{SM2026}.

Having determined the critical coupling $\alphac$, we can now estimate the critical exponents. 
Here we briefly summarize our results, whereas details of our finite-size analysis can be found in the SM \cite{SM2026}.
The scaling form of the correlation ratio,
$R_\mathrm{spin}(L) = f(L^{1/\nu}(\alpha-\alphac),\beta/L^z)$, gives access to the correlation-length exponent $\nu= 0.697(8)$ that is estimated from a crossing analysis of $R_\mathrm{spin}(L)$. 
We use the staggered susceptibility $\chi(q=\pi) \propto L^{2-\eta}$ to obtain the anomalous dimension $\eta=0.01(2)$.
Based on this, we can extract the dynamical critical exponent from the local susceptibility $\chi_\mathrm{loc}  \propto \beta^{(2-d-\eta)/z}$. The latter is plotted in Fig.~\ref{fig:bulkcriticality}(c) for different system sizes and as a function of temperature, for which a power-law fit gives $z=1.96(4)$, thus justifying our choice of $\beta J \propto L^2$ scaling. Furthermore, at $\alpha=\alphac$ the real-space correlations are determined by
$C_\mathrm{spin}(r) \propto r^{-(d+z-2+\eta)}$. 
Figure \ref{fig:bulkcriticality}(d) shows a power-law fit to $C_\mathrm{spin}(r=L/2)$
from which we estimate $\eta + z = 1.99(2)$ or, equivalently, the scaling dimension $\Delta_\mathrm{spin}=0.49(1)$.

Our bulk critical exponents for the dissipative Haldane chain are in excellent agreement with field-theory predictions. We have repeated this analysis for the dissipation-induced transition out of the trivial phase at $\delta=0.4$, for which the resulting critical exponents are again in good agreement, as summarized in Table~\ref{tab:exponents}. As a result, the SPT nature of the Haldane phase does not affect the conventional bulk critical exponents.

\textit{String order parameter.}---%
Until now, we only confirmed that spin excitations remain disordered 
up to a critical coupling $\alphac$, but we have not addressed whether the SPT features of the Haldane phase remain stable towards dissipation. The string order parameter provides one way to test this \cite{PhysRevB.40.4709}. For the isolated chain, 
the string correlation function
$\Cstring(i,j) \vert_{\alpha=0} = \langle S_i^z \big(\prod_{k=i+1}^{j-1} e^{\im \pi S_k^z}\big)S_j^z \rangle$
establishes a finite expectation value for $|i-j|\to \infty$ that
indicates the spontaneous breaking of a hidden $\mathds{Z}_2 \times \mathds{Z}_2$ symmetry within a dual model \cite{PhysRevB.45.304}. It is closely related to the $\mathds{Z}_2 \times \mathds{Z}_2$ dihedral symmetry of $\Hs$ which describes global $\pi$ rotations about the $x$, $y$, and $z$ axes (only two are independent) \cite{Pollmann10}.  For any finite dissipation strength, this symmetry is explicitly broken. However, the global SO(3) symmetry of system plus bath that is generated by the local angular momentum 
$\Jang{i} = \spin{i} + \sum_q \Qbath{iq} \times \Pbath{iq}$ defines an extended dihedral symmetry. Therefore, we can define the generalized string correlation function
\begin{align}
\label{eq:string}
\Cstring(i,j)
    =
    \frac{1}{C_\mathrm{string}^\mathrm{b}(i,j)} \,
    \langle S_i^z \Bigg(\prod_{k=i+1}^{j-1} e^{\im \pi J^z_k} \Bigg) S_j^z \rangle \, .
\end{align}
In the SM \cite{SM2026}, we derive a QMC estimator that allows us to measure $\Cstring(i,j)$ in our simulations.

$C_\mathrm{string}(r)$ displays long-range correlations in the Haldane phase, exponential decay in the AFM phase, and power-law scaling at the critical point. 
In analogy with Eq.~\eqref{eq:Rspin},
we define the string correlation ratio
$\Rstring(L) = 1 - \Cstring(q= \pi+\delta q)/\Cstring(q= \pi)$, where $\Cstring(q)$ is
the Fourier transform of Eq.~\eqref{eq:string}.
Figure \ref{fig:bulkcriticality}(b) shows that
$\Rstring(L)$
scales to one in the Haldane phase and to zero in the AFM phase. 
We determine the critical coupling $\alphac = 0.04350(4)$ and the exponent $\nu=0.77(3)$. Both agree with our previous estimates
from $\Rspin$,
therefore suggesting a direct second-order transition from the Haldane phase to the AFM state. 
At the critical point, $\Cstring(r) \propto r^{- 2 \Dstring}$ decays algebraically,
with a scaling dimension $\Dstring = 0.154(5)$ that is significantly smaller than the one of $\Cspin(r)$ [see Fig.~\ref{fig:bulkcriticality}(d)].

Although the bulk critical exponents are in good agreement with the $\epsilon$ expansion in Eq.~\eqref{eq:Seps},
the nontrivial scaling of $\Cstring(r)$ goes beyond the  bulk theory and is a direct consequence of the SPT nature of the Haldane phase. By contrast, for the transition out of the trivial phase at $\delta=0.4$, $\Rstring(L)$ always scales to zero.

\begin{figure}[t]
  \centering
  \includegraphics[width=\linewidth]{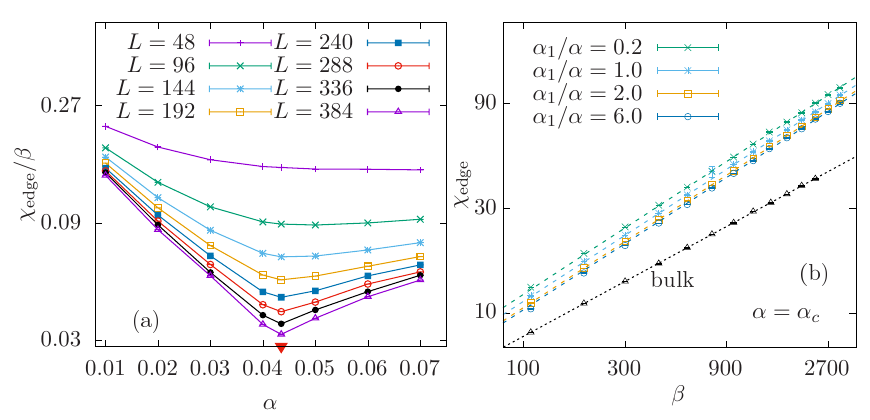}
  \caption{\label{fig:boundaryHaldane}%   
  Boundary criticality of the Haldane chain. (a) Finite-size analysis of the edge local moment, $\chi_\mathrm{edge}/\beta$, as a function of $\alpha$. The red triangle marks the bulk critical coupling $\alphac$. (b) $\chi_\mathrm{edge}(\beta)$ at $\alpha=\alphac$ for different edge couplings $\alpha_1 / \alpha$. For comparison, we include the local susceptibility within the bulk. Dashed lines represent power-law fits.
  }
  \end{figure}

\textit{Boundary criticality.}---%
A characteristic property of the SPT phase is the presence of $S=1/2$ edge states for open boundary conditions. 
Because a free spin exhibits a Curie law with a finite local moment, it can be detected via the local spin susceptibility $\chi_\mathrm{edge} = \chi(i,i)|_{i=1}$ at the edge of our chain. 
Figure \ref{fig:boundaryHaldane}(a) shows that  $\chi_\mathrm{edge}/\beta$ converges to a finite local moment for $\alpha < \alphac$, confirming the stability of the SPT phase. 
We also observe a nonzero local moment for 
$\alpha > \alphac$
which is a direct consequence of symmetry breaking in the AFM phase and even occurs deep in the bulk. Only at $\alpha=\alphac$,  $\chi_\mathrm{edge}/\beta$ vanishes. 

To test how the boundary spin responds to the critical fluctuations of the bulk critical point, we plot $\chi_\mathrm{edge}$ as a function of $\beta$ in Fig.~\ref{fig:boundaryHaldane}(b). We find that the local susceptibility diverges faster at the boundary than in the bulk. By fitting $\chi_\mathrm{edge} \propto \beta^{(2-d-\eta_\parallel)/z}$, we estimate the boundary anomalous dimension $\eta_\parallel = -0.28(1)$. We notice that $\eta_\parallel$ remains constant even if we tune the boundary dissipation strength $\alpha_1 / \alpha$ away from unity.

To classify the surface criticality correspondingly, we have extended the $\epsilon$ expansion of Eq.~\eqref{eq:Seps} to a semi-infinite chain with ordinary boundary conditions. Details are given in the SM \cite{SM2026}. For $N=3$ we obtain $\eta_\parallel = 1.288$.
Although our bulk anomalous dimension  is in good agreement with the $\epsilon$ expansion, $\eta_\parallel$ differs significantly. Therefore, this transition falls into a distinct boundary universality class. The underlying reason is that the dissipative O(3) theory does not include the topological term $\Stop$ in open boundary condition.  The latter  is important to capture the edge states of the Haldane  phase as well as the surface criticality at the bulk transition point.  

\begin{figure}[t]
  \centering
  \includegraphics[width=\linewidth]{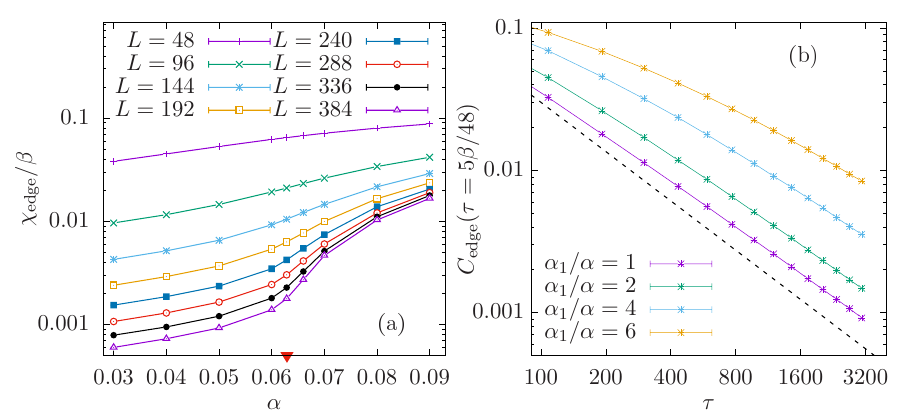}
  \caption{\label{fig:boundaryTrivial}%
  Boundary response of the dimerized chain. (a) Finite-size analysis of the edge local moment, $\chi_\mathrm{edge}/\beta$,  as a function of $\alpha$. The red triangle marks the bulk critical coupling $\alphac$. (b) $C_\mathrm{edge}(\tau = 5\beta / 48)$ at  $\alpha=\alphac$  for different edge couplings $\alpha_1/\alpha$. The black dashed line denotes the $\epsilon$ expansion result. 
  }
  \end{figure}

To test the applicability of the dissipative O(3) boundary theory, 
we repeat our analysis for the dissipation-induced transition out of the trivial phase at $\delta=0.4$ 
\footnote{In Eq.~\eqref{eq:def_dimerized} we made a specific choice of boundary condition. Reversing the sign of $\delta$ leaves an unpaired $S=1$ degree of freedom at the boundary. The role of dangling edge states for boundary criticality without symmetry protection is still under debate. Here we only focus on the \textit{trivial} case of $\delta>0$.}. 
Now the local moment $\chi_\mathrm{edge}/\beta$ scales to zero in the disordered phase, as shown in Fig.~\ref{fig:boundaryTrivial}(a). To determine the boundary anomalous dimension, we calculate the time-displaced correlation function $C_\mathrm{edge}(\tau) = C(r=0;\tau)$ at the edge, which we  show in Fig.~\ref{fig:boundaryTrivial}(b) for fixed $\tau=5\beta/48$. At criticality, a fit to the power-law form $C_\mathrm{edge}(\tau) \propto \tau^{-(d+z-2+\eta_\parallel)/z}$ gives $\eta_\parallel = 1.27(3)$. Again, the boundary anomalous exponent is robust towards increasing the boundary dissipation strength $\alpha_1/\alpha$, but finite-size corrections become stronger. All in all, the boundary criticality out of the trivial phase is in good agreement with the dissipative O(3) theory due to the absence of  boundary Berry phase. 

We note that the correlations between edge and bulk are described by an additional exponent $\eta_\perp = (\eta + \eta_\parallel)/2$ which, however, is not independent. A detailed analysis for $\eta_\perp$ is included in the SM \cite{SM2026} where we also discuss
direct simulations of the dissipative nonlinear sigma model.

\textit{Spin defects in critical antiferromagnets.}---%
The boundary of our dissipative Haldane chain realizes the rich phenomenology of Bose Kondo impurity models.
Weak coupling between the $S=1/2$ boundary mode and our \textit{noninteracting} bosonic bath yields the SU(2)-symmetric spin-boson model for which ohmic dissipation constitutes a marginally irrelevant perturbation \cite{PhysRevB.61.4041}; 
the Curie law observed in Fig.~\ref{fig:boundaryHaldane}(a) might therefore be affected by logarithmic corrections \cite{PhysRevLett.130.186701}.
At $\alpha=\alpha_\mathrm{c}$,
our boundary resembles an impurity in a critical antiferromagnet \cite{SachdevImpurity99, Vojta_2000}, for which
the bulk critical fluctuations provide an \textit{interacting} environment. Interaction effects are a relevant perturbation \cite{SachdevImpurity99, Vojta_2000}
that induces nontrivial boundary criticality and have been studied numerically in (2+1)D quantum magnets \cite{PhysRevLett.98.087203, PhysRevB.70.024406}. Recently, the Bose Kondo problem has gained renewed interest in the context of boundary conformal field theories \cite{Cuomo_2022, 2025arXiv250814963K, 2026arXiv260407554S}.
In our case, the nontrivial boundary criticality is driven by a \textit{nonconformal} bulk theory.
In contrast to many other realizations of boundary criticality, the effective dimension of our dissipative bulk theory is not fixed but can be tuned continuously via the bath exponent $s$. For the dissipative O(2) chain,  the tunability of bulk exponents
was studied for $1 \leq s \leq 2$ using classical Monte Carlo simulations \cite{PhysRevB.85.214302}, whereas the boundary response has remained unexplored.
Our dissipative Haldane chain opens up a route for tuning nontrivial boundary criticality. 
In particular for $s<1$, the free edge spin might become unstable towards a critical intermediate-coupling fixed point, as described by the SU(2)-symmetric spin-boson model \cite{PhysRevB.61.4041, Vojta_2000}, which can experience fixed-point annihilation and even long-range order \cite{PhysRevLett.130.186701, Weber2025}. While these features would apply to the boundary of the disordered phase, it raises the question if nontrivial boundary effects can even occur at a potential 
bulk critical point for $s<1$.
Interestingly, fractionalized quasiparticles were reported in related setups
\cite{PhysRevA.96.053620, PhysRevB.98.214516}.

\textit{Conclusions.}---%
We have explored the role of the SPT phase
in dissipation-induced quantum criticality by comparing the Haldane chain with its topologically trivial counterpart. 
Although both models share the same bulk universality class, the SPT character of the Haldane chain remains detectable through a generalized string order parameter, which acquires a nontrivial scaling dimension at the critical point.
At the boundary, however, the distinction between the two systems becomes manifest. While the trivial chain exhibits ordinary boundary criticality described by the $\epsilon$ expansion, the protected edge states of the Haldane chain induce a distinct boundary universality class.
In future work, it will be interesting to explore how the nonconformal nature of our dissipation-induced bulk criticality can lead to novel phenomena beyond the framework of boundary conformal field theory. In particular, its high controllability through the effective bath dimension $s$ gives access to tunable boundary criticality beyond fixed spatial dimensions. It will be worth to investigate whether the rich phenomenology of three-dimensional O($N$) models, with their various boundary universality classes \cite{Toldin_2021,   Metlitski_2022, PhysRevLett.128.215701}, can be realized or even modified in dissipative O($N$) models. Moreover, it has recently been shown that even critical bulk theories can be enriched with SPT properties \cite{PhysRevX.7.041048, PhysRevX.11.041059}; it remains open if this can also occur in open quantum systems.

\begin{acknowledgments}
\textit{Acknowledgments.}---%
We thank F.~Assaad, H.W.~Diehl, W.~Guo, L.~Janssen, S.~Sachdev, and L.~Zhang for helpful discussions. 
Z.W. was supported by the National Natural Science Foundation of China under Grant No. 12504273. 
M.W. was supported by the Deutsche Forschungsgemeinschaft
through the W\"urzburg-Dresden Cluster of Excellence \textit{ctd.qmat}---Complexity, Topology, and Dynamics in Quantum Matter (EXC 2147, Project No. 390858490).
M.W. was also supported by France 2030 funding: ANR-21-EXES-0003.
\end{acknowledgments}

\bibliography{./zwang}

\clearpage

\stepcounter{myequation}
\stepcounter{myfigure}
\stepcounter{mytable}

\setcounter{secnumdepth}{3}  % add numeration of sections

\renewcommand{\thefigure}{S\arabic{figure}}
\renewcommand{\thesection}{S\arabic{section}}
\renewcommand{\thetable}{S\arabic{table}}
\renewcommand{\theequation}{S\arabic{equation}}

\onecolumngrid

\title{Dissipation-induced bulk and boundary criticality in the Haldane chain}

\centerline{\bf\large Supplemental Material} \vskip3mm
\centerline{\bf\large for} \vskip3mm
\centerline{\bf\large Dissipation-induced bulk and boundary criticality in the Haldane chain} \vskip1cm

\maketitle

\twocolumngrid

\section{Quantum Monte Carlo method and generalized string operator}
\label{Sec:QMC}

The goal of this section is to derive a QMC estimator for the generalized string operator. As this requires some background on our QMC method and its formulation for retarded interactions, we will first review the necessary steps following Ref.~\cite{PhysRevB.105.165129}.

\subsection{Interaction representation and retarded interactions}

Our QMC method is formulated in the interaction picture where we split the full Hamiltonian
\begin{align}
H= H_0 + V
\end{align}
into the noninteracting part
\begin{align}
H_0 = H_\mathrm{b} = \sum_{ iq } \omega_{q} \, \acr{iq} 
  \cdot \aan{iq}
\end{align}
given by the free bosonic bath and the remainder
\begin{align}
V = \Hs + \Hsb \, , \quad
\Hsb =  \sum_{iq} \lambda_{q} \, \big( \acr{iq} + \aan{iq} \big)  \cdot \spin{i} \, .
\end{align}
As usual, we define the time-dependent operators
$V(\tau) = e^{\tau H_0} V e^{-\tau H_0}$ 
and the expectation value with respect to the noninteracting part
$\langle O \rangle_0 = \frac{1}{Z_0} \Tr e^{-\beta H_0} O$. Note that the trace runs over spin and bosonic degrees of freedom.

The trace over the bosons can be calculated exactly so that the partition function of the full system \cite{PhysRevB.105.165129}
\begin{align}
\nonumber
Z
&=Z_0 \, \langle \Ttime \, e^{-\int d\tau \,V(\tau)}\rangle_0
\\
&= Z_\mathrm{b} \Trs \Ttime \, e^{-\Hfullint}  \, ,
\qquad
\Hfullint = \Hsint + \Hret \, ,
\label{eq:Zint}
\end{align}
reduces to a trace over the spin degrees of freedom which are included in the effective spin interaction $\Hfullint$.
Here, we have defined
$\Hsint = \int d\tau \, \Hs(\tau)$.
Note that the time evolution of spin operators with respect to $H_0$ is trivial, but the time labels are required to perform the time ordering.
The spin-boson interaction is replaced by a retarded spin-spin interaction
\begin{align}
\Hret = - \iint_0^\beta  d \tau \, d \tau' \sum_i  K( \tau - \tau' ) \,  \spin{i}(\tau)
\cdot \spin{i}(\tau') 
\end{align}
that is mediated by the free bath propagator
\begin{align}
K(\tau) =
\frac{1}{\pi}
\int_0^{\infty} d\omega \, J(\omega) \,
D(\omega,\tau)
\, .
\end{align}
Here we have taken the continuum limit of the bath to include the bath spectral function $J(\omega)$ as defined in the main paper.
For a single frequency, the boson propagator is given by
\begin{align}
D(\omega,\tau)
=
\frac{e^{-\omega \tau}}{1-e^{-\beta \omega}}
\, , \quad
D(\omega,\tau+\beta) = D(\omega,\tau)
\, .
\end{align}
For $J(\omega) \propto \omega^s$, the full bath propagator exhibits a power-law decay $K(\tau) \propto 1/\tau^{1+s}$ for $\omc \tau \gg 1$ where $\omc$ is the cutoff frequency of the bath.

\subsection{Basics of our QMC method}

\subsubsection{Configuration space and Monte Carlo weights}

Our QMC method \cite{Weber17, PhysRevB.105.165129} is based on a perturbation expansion of the partition function in Eq.~\eqref{eq:Zint}. Because we expand in the full exponent $\Hfullint$, this is in close analogy to the stochastic series expansion which can be reformulated in the interaction representation \cite{PhysRevB.56.14510}.
We obtain
\begin{align}
\label{eq:ZMC_conf1}
\frac{Z}{\Zb}
&=
\sum_{n=0}^\infty \frac{(-1)^n}{n!} \sum_{\conf_n}
\sum_\alpha
\braket{\alpha|\Ttime \, \mathcal{H}_{\nu_n} \dots \mathcal{H}_{\nu_1}|\alpha}
\, .
\end{align}
The trace over spin states is performed in the $S^z$ eigenbasis $\ket{\alpha}$, $n$ is the expansion order, and
$\conf_n = \{ \nu_1, \dots, \nu_n \}$ is a list of internal vertex variables which will be specified below. With this, a Monte Carlo configuration is defined by $\conf = \{n, \conf_n, \alpha\}$ and the Monte Carlo weight $W(\conf)$ can be identified from the partition function
\begin{align}
\frac{Z}{\Zb}
=
\sum_{\conf} W(\conf) \, .
\end{align}
As in the stochastic series expansion, the trace over all operators in Eq.~\eqref{eq:ZMC_conf1} factorizes into individual vertex weights $w_\nu$, but only after performing an explicit time ordering \cite{Weber17}. We obtain
\begin{align}
W(\conf) = \frac{(-1)^n}{n!} \prod_{k=1}^n W_{\nu_k} \, .
\end{align}
The perturbation expansion can be sampled efficiently using the diagonal and directed-loop updates developed for the stochastic series expansion \cite{Sandvik99b, Syljuasen02} which generalize to our case of retarded interactions \cite{Weber17, PhysRevB.105.165129}. For details, we refer to the existing literature. In the following, we only define the interaction vertices necessary to calculate the string order parameter.

\subsubsection{Interaction vertices}
\label{Sec:VertInt}

Because $\Hfullint$ consists of two distinct interactions, we also need to define two types of vertices. Therefore, we redefine the vertex variables as $\nu = \{ \Gamma, \nu_\Gamma \}$. Here, $\Gamma$ distinguishes between Heisenberg and retarded interactions and $\nu_\Gamma$ includes the internal vertex variables which differ for the two interactions.

For the Heisenberg interaction, the vertex variables are $\nu_\mathrm{H} = \{i, \tau, a\}$ and the local weights 
$W_{\nu_k}$
are defined as expectation values of the propagated states $\ket{\alpha_k}$. For the off-diagonal operators, we have
\begin{align}
W_{i,\tau,a=1} = \frac{J}{2} \braket{\alpha_k |( S_i^+ S_{i+1}^- + S_{i+1}^+ S_{i}^- ) | \alpha_{k-1}}
\end{align}
labeled by $a=1$ and for the diagonal ones
\begin{align}
W_{i,\tau,a=2} = - J \braket{\alpha_k | ( C - S_i^z S_{i+1}^z ) | \alpha_{k-1}}
\end{align}
labeled by $a=2$. A constant shift $C$ to the Hamiltonian does not change the physics, but
avoids a sign problem in the diagonal weights $(-W_{i,\tau,a=2}) \geq 0$. Moreover, on a bipartite lattice, the off-diagonal weights $(-W_{i,\tau,a=1})$ do not lead to a sign problem because they always appear in pairs to close the trace.

To implement the $S=1$ operators in our algorithm, we split
$\spin{i} = \spinsplit{i1} + \spinsplit{i2}$
into a sum of two $S=1/2$ operators \cite{PhysRevLett.73.1295}.
Instead of fixing the  
$S=1$ Hilbert space by manually projecting out the spin singlet state, we impose this constraint energetically. This is possible because $( \spinsplit{i1} + \spinsplit{i2})^2$ is a conserved quantity at each site and the singlet gap is an ultraviolet energy scale of $\mathcal{O}(J)$ for all system sizes. We confirmed numerically that $\beta J \approx 3$ is sufficient to satisfy the $S=1$ constraint.

For the retarded spin interaction, the vertex variables are $\nu_\mathrm{ret} = \{i, \omega, \tau, \tau', a\}$
and the vertex weights are given by
$W_{\nu_{k}} = - J(\omega_{k}) \, D(\omega_k, \tau_k - \tau'_k) \, \tilde{W}_{\nu_k}$. The off-diagonal weights are
\begin{align}
\nonumber
\tilde{W}_{i,\omega,\tau,\tau',a=1}
=
\frac{1}{2} & \left[ \braket{\alpha_k | S_i^+ |\alpha_{k-1}}
\braket{\alpha'_k | S_i^- |\alpha'_{k-1}} \right.
\\
+
& \phantom{[}\left. \braket{\alpha_k | S_i^- |\alpha_{k-1}}
\braket{\alpha'_k | S_i^+ |\alpha'_{k-1}} \right]
%\, .
\label{eq:weight_Vret}
\end{align}
and the diagonal weights
\begin{align}
\tilde{W}_{i,\omega,\tau,\tau',a=2}
=
C + \braket{\alpha_k | S_i^z |\alpha_{k-1}}
\braket{\alpha'_k | S_i^z |\alpha'_{k-1}} \, .
\end{align}
Here, the propagated states $\ket{\alpha_k}$ and $\ket{\alpha'_k}$ correspond to different imaginary times $\tau$ and $\tau'$, respectively. The weights $\tilde{W}_\nu$ are equivalent to the ones of the ferromagnetic Heisenberg model, therefore the split-spin representation for $S=1$ can be applied as before. 
The solution of the directed-loop equations works as for the Heisenberg model and leads to the nonlocal wormhole moves \cite{PhysRevB.105.165129}. The global prefactors to the weight, \ie, $J(\omega)$ and $D(\omega, \tau-\tau')$ can be sampled efficiently during the diagonal updates. For further details, see Ref.~\cite{PhysRevB.105.165129}.

\subsection{QMC estimator for the string operator}

\subsubsection{Definition of the string operator}

For the isolated Haldane chain, the string operator is closely related to the $\mathds{Z}_2 \times \mathds{Z}_2$ dihedral symmetry of global $\pi$ rotations about the spin $x$, $y$, and $z$ axes (but only two of them are independent). This symmetry is automatically conserved 
in the presence of an SO(3) spin rotational symmetry, but it is broken for any finite coupling to the environment.

Our quantum dissipative spin chain conserves total angular momentum of system plus bath which is generated by the local operators 
\begin{align}
\Jang{i} = \spin{i} + \Lang{i} \, ,
\qquad
\Lang{i} = \sum_q \Qbath{iq} \times \Pbath{iq} \, .
\end{align}
Here, $\spin{i}$ is the local $S=1$ operator
and $\Lang{i}$ the angular momentum operator of a local bosonic bath.
The latter is defined via the position and momentum operators
\begin{align}
\Qbath{iq} &= \frac{1}{\sqrt{2M_q \omega_q}} \big( \acr{iq} + \aan{iq} \big) \, ,
\\
\Pbath{iq} &= \im \sqrt{\frac{M_q \omega_q}{2}} \big( \acr{iq} - \aan{iq} \big) \, .
\end{align}
The SO(3) symmetry of system plus bath contains an extended dihedral symmetry so that we can define the generalized string operator
\begin{align}
\label{eq:AppString}
\Cstring(i,j)
    =
    \frac{1}{C_\mathrm{string}^\mathrm{b}(i,j)}
    \langle S_i^z \Bigg(\prod_{k=i+1}^{j-1} e^{\im \pi J^z_k} \Bigg) S_j^z \rangle
\end{align}
via the local $\pi$ rotations of system plus bath. To recover the definition of the isolated system for $\alpha=0$, we have normalized the string operator by an additional factor of $C_\mathrm{string}^\mathrm{b}(i,j) = \langle \prod_{k=i+1}^{j-1} e^{\im \pi L^z_k} \rangle_\mathrm{b}$ that is calculated with respect to the noninteracting bosonic bath.

Note that we define the string operator via the $z$ component of the angular momentum operators because our QMC simulations are performed in the $S_i^z$ eigenbasis. This simplifies the calculation of the spin contribution to $C_\mathrm{string}(i,j)$. In the following, we derive how to recover the bosonic part of $C_\mathrm{string}(i,j)$.

\subsubsection{Basis transformation}
\label{Sec:App_BasisTrafo}

In second quantization, the $z$ component of the bath angular momentum operator takes the form
\begin{align}
\nonumber
L_i^z &= - \im \sum_q \big( \acrc{iqx} \aanc{iqy} - \acrc{iqy} \aanc{iqx} \big)
\\
&= - \sum_q \big( \bcrc{iq1} \banc{iq1} - \bcrc{iq2} \banc{iq2} \big)
\, .
\end{align}
In the last step, we have used the basis transformation
\begin{equation}
\begin{pmatrix}
  \banc{iq1} \\ \banc{iq2}     
\end{pmatrix} 
= \frac{1}{ \sqrt{2} } 
 \begin{pmatrix}
     1 & \im \\  1 & -\im 
 \end{pmatrix} 
 \begin{pmatrix}
    \aanc{iqx} \\ \aanc{iqy}
 \end{pmatrix} \, ,
 \quad \banc{iq3} = \aanc{iqz}
\end{equation}
to diagonalize $L_i^z$. With this, the different contributions to our spin-bath Hamiltonian become
\begin{gather}
H_\mathrm{b} = \sum_{iq\ell} \omega_q \bcrc{iq\ell} \banc{iq\ell} \, ,
\\ \nonumber
\Hsb=  \sum_{iq} \lambda_q \, \Big[ \big( \bcrc{iq3} + \banc{iq3} \big) \, S_i^z
\qquad\qquad \\
+ \frac{1}{\sqrt{2}}\big( \bcrc{iq1} + \banc{iq2} \big) \, S_i^+
+ \frac{1}{\sqrt{2}} \big( \bcrc{iq2} + \banc{iq1} \big) \, S_i^- \Big] \, .
\end{gather}
In particular, the bosonic contribution to the string order parameter reduces to the calculation of the local parity operator
$
\mathcal{P}_{i}^\mathrm{b}
= e^{\im \pi L_i^z}
$.

\begin{table*}[t]
\caption{%
\label{tab:CompareParity}%
Comparison of the main equations with and without parity operator.
}
\begin{ruledtabular}
\renewcommand{\arraystretch}{2.5} 
\begin{tabular}{l|c|c}
  & Standard formalism  & With parity operator $\displaystyle\mathcal{P}_\mathrm{b} = e^{\im \pi \sum_\mu \bcrc{\mu} \banc{\mu}}$ \\
\colrule
Hamiltonian &
\multicolumn{2}{c}{%
$
\displaystyle
H = H_\mathrm{b} + H_\mathrm{sb}
$
\, ,
\qquad
$
\displaystyle
H_\mathrm{b} = \sum_\mu \omega_\mu \bcrc{\mu} \banc{\mu}
$ \, ,
\qquad
$
\displaystyle
H_\mathrm{sb} = \sum_\mu \big( \bcrc{\mu} \vrho{\mu} + \vrhodag{\mu} \banc{\mu}  \big)
$
}
\\
\colrule
Expectation value
&
$
\displaystyle
\langle O \rangle_\mathrm{b} = \frac{1}{\Zb} \Trb e^{-\beta H_\mathrm{b}} O
$
&
$
\displaystyle
\langle O \rangle_\mathrm{b}^\mathcal{P} = \frac{1}{\Zb^\mathcal{P}} \Trb e^{-\beta H_\mathrm{b}} \mathcal{P}_\mathrm{b} \, O
$
\\
Partition function
& 
$
\displaystyle
\Zb = \Trb e^{-\beta H_\mathrm{b}}
= \prod_\mu \frac{1}{1- e^{-\beta \omega_\mu}}
$
& 
$
\displaystyle
\Zb^\mathcal{P} = \Trb e^{-\beta H_\mathrm{b}} \mathcal{P}_\mathrm{b}
= \prod_\mu \frac{1}{1+e^{-\beta \omega_\mu}}
$
\\
Boson propagator
&
$
\displaystyle
D(\omega_\mu,\tau-\tau')
= \langle \Ttime \banc{\mu}(\tau) \bcrc{\mu}(\tau') \rangle_\mathrm{b}
$
&
$
\displaystyle
D^\mathcal{P}(\omega_\mu,\tau-\tau')
= \langle \Ttime \banc{\mu}(\tau) \bcrc{\mu}(\tau') \rangle_\mathrm{b}^\mathcal{P}
$
\\
&
$
\displaystyle
D(\omega,\tau)
= 
\frac{e^{-\omega \tau}}{1-e^{-\beta \omega}}
$
\quad for \quad
$0 \leq \tau < \beta$
&
$
\displaystyle
D^\mathcal{P}(\omega,\tau)
= 
\frac{e^{-\omega \tau}}{1+e^{-\beta \omega}}
$
\quad for \quad
$0 \leq \tau < \beta$
\\
&
$
\displaystyle
D(\omega,\tau+\beta) = D(\omega,\tau)
$
&
$
\displaystyle
D^\mathcal{P}(\omega,\tau+\beta) = - D^\mathcal{P}(\omega,\tau)
$
\\
Trace over bosons
&
$
\displaystyle
\langle \Ttime \, e^{-\int d\tau \, H_\mathrm{sb}(\tau)} \rangle_\mathrm{b}
=
\Ttime \, e^{-\Hret}
$
&
$
\displaystyle
\langle \Ttime \, e^{-\int d\tau \, H_\mathrm{sb}(\tau)} \rangle_\mathrm{b}^\mathcal{P}
=
\Ttime \, e^{-\Hret^\mathcal{P}}
$
\\
Retarded interaction
&
$
\displaystyle
\Hret = - \iint d\tau \, d\tau'  \sum_\mu \vrhodag{\mu}(\tau) \, D(\omega_\mu, \tau - \tau') \, \vrho{\mu}(\tau')
$
&
$
\displaystyle
\Hret^\mathcal{P}
=
- \iint d\tau \, d\tau'  \sum_\mu \vrhodag{\mu}(\tau) \, D^\mathcal{P}(\omega_\mu, \tau - \tau') \, \vrho{\mu}(\tau')
$
\end{tabular}
\end{ruledtabular}
\end{table*}

\subsubsection{Bosonic parity operator}

To keep our derivation of the parity operator as generic as possible, we use the generalized representation of the spin-boson Hamiltonian $H = H_\mathrm{b} + H_\mathrm{sb}$ where
\begin{align}
H_\mathrm{b} = \sum_\mu \omega_\mu \bcrc{\mu} \banc{\mu} \, ,
\quad
H_\mathrm{sb} = \sum_\mu \big( \bcrc{\mu} \vrho{\mu} + \vrhodag{\mu} \banc{\mu} \big)
\, .
\end{align}
Here we introduce a superindex $\mu$ for the bath modes and couple the bosons to a generic spin operator $\varrho_\mu$. We drop the pure spin part $\Hs$ to simplify our notation.
The steps to derive a retarded interaction within this notation are the same as before and have been summarized in the left column of Table~\ref{tab:CompareParity}.

To include measurements of the parity operator
\begin{align}
\mathcal{P}_\mathrm{b} = e^{\im \pi \sum_\mu \bcrc{\mu} \banc{\mu}}
\end{align}
in our formalism, we only need to realize that
\begin{align}
e^{-\beta H_\mathrm{b}} \mathcal{P}_\mathrm{b} 
=
e^{-\beta H_\mathrm{b}^{\mathcal{P}}}
\end{align}
defines an effective noninteracting bath Hamiltonian
\begin{align}
H_\mathrm{b}^{\mathcal{P}} = \sum_\mu \left(\omega_\mu - \frac{\im \pi} {\beta}\right) \bcrc{\mu} \banc{\mu} \, .
\end{align}
Therefore, instead of keeping $\mathcal{P}_\mathrm{b}$ as part of the observable, we can redefine all expectation values
\begin{align}
\langle \mathcal{P}_\mathrm{b} O \rangle_\mathrm{b}
=
\frac{\Zb^\mathcal{P}}{\Zb} \langle O \rangle_\mathrm{b}^\mathcal{P}
\end{align}
in terms of the parity expectation value $\langle O \rangle_\mathrm{b}^\mathcal{P}$ defined in the right column of Table~\ref{tab:CompareParity}. 
The entire formalism of integrating out the bosonic bath can be easily transferred to the parity formalism which is summarized in the right column of Table~\ref{tab:CompareParity}.
The main change is that we need to substitute $e^{-\beta \omega_\mu} \to - e^{-\beta \omega_\mu}$ in the partition function and in the bath propagator, effectively modifying the exchange  statistics. Thereby, the bath propagator also becomes antisymmetric under time translations $\tau \to \tau + \beta$. Note that the time evolution of, \eg, $\banctext{\mu}(\tau)$ is not affected by this.
Moreover, Wick's theorem remains valid so that, after tracing out the bath, we arrive at the same retarded interaction in which only the boson propagator is replaced.

The similarity of the parity formalism with the original one suggests that, within our QMC method, we can calculate the parity operator by reweighting the interaction expansion of the retarded interaction. Therefore, we first need to reformulate the original interaction expansion in a way that is in accordance with our Monte Carlo sampling. We write the partition function
\begin{align}
\frac{Z}{\Zb} 
= 
\langle \Ttime \, e^{- \Hret} \rangle_\mathrm{s}
=
\sum_\conf W(\conf)
\end{align}
as a sum over configurations $\conf = \{n, \conf_n\}$ where $n$ is the expansion order and $\conf_n = \{ \nu_1, \dots, \nu_n \}$ is a list of vertices with variables $\nu = \{ \mu, \tau, \tau' \}$. The weight reads
\begin{align}
W(\conf) = \frac{w(\conf)}{n!}
\prod_{k=1}^n D(\omega_{\mu_k},\tau_k - \tau'_k)
\end{align}
and we define the operator weight
\begin{align}
w(\conf) = 
\langle \Ttime \, \prod_{k=1}^n \vrhodag{\mu_k}(\tau_k)  \, \vrho{\mu_k}(\tau'_k) \rangle_\mathrm{s}
\, .
\end{align}
Note that, for our derivation, it is not necessary to split $w(\conf)$ into individual vertex weights, as it is done to arrive at the stochastic-series-expansion representation. Therefore, we also do not introduce the state $\ket{\alpha}$ in our configurations.
Moreover, expectation values of operators
\begin{align}
\langle O \rangle
=
\frac{1}{Z} \sum_\conf W(\conf) \, \langle O \rangle_\conf
\end{align}
can now be calculated per configuration $\conf$ via $\langle O \rangle_\conf$.

For the parity formalism, the weights $W(C)$ only differ by the product of boson propagators whereas the operator weights $w(\conf)$ are the same. Therefore, the expectation value of the parity operator can be easily recovered from the weight $W(\conf)$ of the original formalism. We get
\begin{align}
\frac{\langle \mathcal{P}_\mathrm{b} \rangle_\conf}
{\langle \mathcal{P}_\mathrm{b} \rangle_\mathrm{b}}
=
\prod_{k=1}^n \frac{D^\mathcal{P}(\omega_{\mu_k}, \tau_k - \tau'_k)}{D(\omega_{\mu_k}, \tau_k - \tau'_k)} \, .
\end{align}
For each vertex, we define the ratio of absolute values
\begin{align}
\label{eq:Gom}
G(\omega)
= 
\left|
\frac{D^\mathcal{P}(\omega, \tau)}{D(\omega, \tau)}
\right|
=
\frac{1-e^{-\beta \omega}}{1+e^{-\beta \omega}}
\end{align}
and their sign %for $0 \leq \tau < \beta$
\begin{align}
\label{eq:Ftau}
F(\tau)
=
\mathrm{sgn} \left[ \frac{D^\mathcal{P}(\omega, \tau)}{D(\omega, \tau)} \right]
=
\begin{cases}
+1 \, , \quad \phantom{-}0 \leq \tau < \beta\\
-1 \, , \quad -\beta \leq \tau < 0
\end{cases}
\, .
\end{align}
Eventually, we arrive at the final estimator for the parity operator which becomes
\begin{align}
\label{eq:QMC_parity}
\frac{\langle \mathcal{P}_\mathrm{b} \rangle_\conf}
{\langle \mathcal{P}_\mathrm{b} \rangle_\mathrm{b}}
=
\prod_{k=1}^n G(\omega_{\mu_k}) \, F(\tau_k - \tau'_k) \, .
\end{align}
Hence, the two functions need to be evaluated for all retarded vertices at the values of the vertex variables.
Note that our derivation only required basic properties of the interaction expansion and therefore applies to other continuous-time QMC methods that are based on a similar formulation \cite{Gull_rev}.

\subsubsection{Implementation and testing}

Now that we know how to calculate the bosonic parity operator, we can return to the string correlation function defined in Eq.~\eqref{eq:AppString}. For a given Monte Carlo configuration $\conf=\{n,\conf_n,\alpha\}$, the spin expectation values can be easily calculated from the state $\ket{\alpha}$ which is written in the $S^z$ eigenbasis. Additionally, we need to calculate the expectation values of the local bosonic parity operators $e^{\im \pi L_i^z}$. From the basis transformation in Sec.~\ref{Sec:App_BasisTrafo},
it becomes clear that they are directly related to the nonlocal spin flips in time and therefore need to be recovered from the off-diagonal ($a=1$) retarded vertices defined in Sec.~\ref{Sec:VertInt}.
We calculate $G(\omega_k)$ according to Eq.~\eqref{eq:Gom} using the frequency variable $\omega_k$ of the retarded spin-flip vertex, whereas $F(\tau_k -\tau'_k)$ in Eq.~\eqref{eq:Ftau} only requires us to compare which of the two time variables of the vertex is larger. To estimate the local parity operator according to Eq.~\eqref{eq:QMC_parity}, we only need to multiply this information for every spin-flip vertex on the corresponding lattice site. Note that the normalization $C^\mathrm{b}_\mathrm{string}(i,j)$ is automatically taken into account because it corresponds to $\langle\mathcal{P}_\mathrm{b}\rangle_\mathrm{b}$ in Eq.~\eqref{eq:QMC_parity}.

Our QMC estimator for the boson parity in Eq.~\eqref{eq:QMC_parity} requires a reweighting of all spin-flip vertices which shows up as a product over vertices. In particular, each individual factor $G(\omega_k)$ can cover a range between 0 and 1 depending on whether $\beta \omega_k \ll 1$ or $\beta \omega_k \gg 1$, respectively, which can substantially affect the local contribution to $C_\mathrm{string}(i,j)$. In the following, we want to test our QMC estimator of $C_\mathrm{string}(i,j)$ for different scenarios.

\begin{figure}
  \centering
\includegraphics[width=0.8\linewidth]{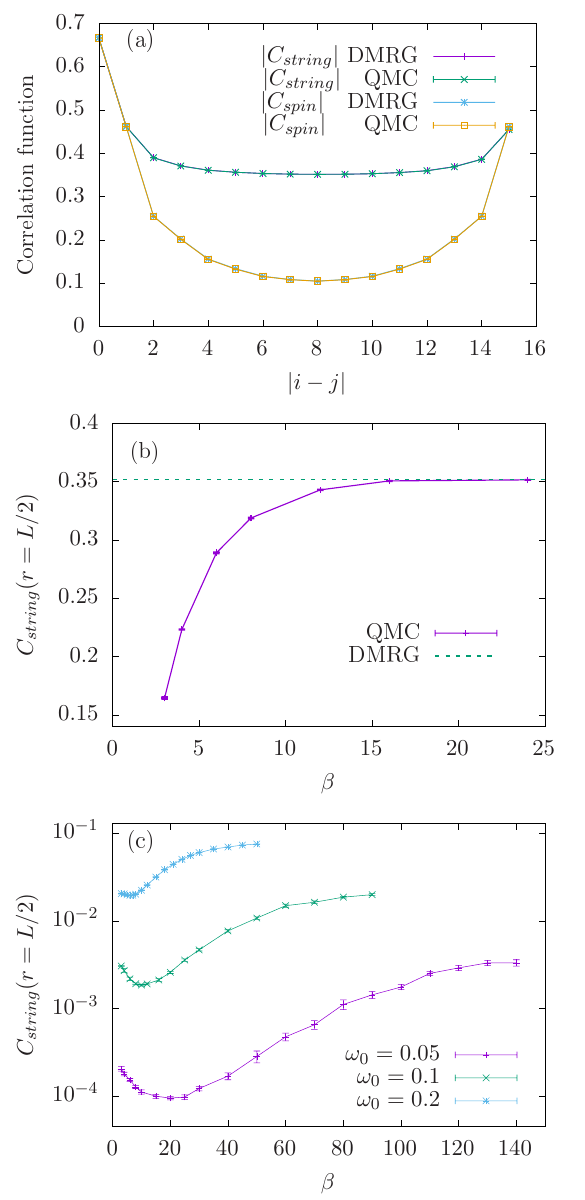}
\caption{\label{fig:dmrg}%
(a) Benchmark 
of our QMC estimators for the spin and string correlation functions with DMRG. Here we use a single boson frequency $\omega_0 / J=1$, a coupling of $\lambda=0.2$, $L=16$ sites, and an inverse temperature of $\beta J =48$.
(b) Convergence of $ C_{\text{string} }(r = L/2 ) $ with inverse temperature $\beta$ for the same parameters.
The dashed line represents the DMRG result for zero temperature. 
(c) $ C_{\text{string} }( r = L/2 ) $ as function of $\beta$ for 
$\omega_0 / J \in \{0.05, 0.1,0.2\}$ and the remaining parameters set as before. 
}
\end{figure}

First, we consider a simplified spin-boson coupling
\begin{equation}
 H_\mathrm{s} + \Hsb = \omega_{0}  \sum_{ i } \acr{i} 
  \cdot \aan{i} + 
  \lambda \sum_{i} \big( \acr{i} + \aan{i} \big)  \cdot \spin{i}
\label{Eq:def_Hsb}  
\end{equation}
where the continuous bath is replaced by a single boson frequency $\omega_0$. This has the advantage that we can compare our results with the density-matrix renormalization group (DMRG). Figure \ref{fig:dmrg}(a) shows such a comparison for the real-space spin and string correlations as a function of distance, which are in excellent agreement with each other.
Here we use a periodic chain of $L=16$ sites and model parameters $\omega_0/J =1$ and $\lambda/J = 0.2$.
In our DMRG calculation, we use a Hilbert-space truncation of maximum $10$ bosons per site and a bond dimension of $\chi=200$ which is sufficient to ensure convergence. Our calculation is based on the ITensor library \cite{10.21468/SciPostPhysCodeb.4}.
Note that 
the normalization factor $C^\mathrm{b}_\mathrm{string}(i,j)$ 
converges to one for
$\beta\to\infty$.
To further test the temperature convergence of our QMC results, we show $C_\mathrm{string}(r=L/2)$ in Fig.~\ref{fig:dmrg}(b) as a function of inverse temperature $\beta$. We confirm that our QMC estimator converges to the DMRG prediction when $\beta$ is large enough. Finally, we also want to test how the temperature convergence of our estimator depends on the boson frequency $\omega_0$. To this end, we compare $C_\mathrm{string}(r=L/2)$ for $\omega_0 /J \in\{0.05, 0.1, 0.2\}$ in Fig.~\ref{fig:dmrg}(c). Indeed, we need higher $\beta$ to achieve temperature convergence at lower $\omega_0$. This is expected from our QMC estimator in Eq.~\eqref{eq:QMC_parity}, because only for $\beta \omega_0 \gg 1$ we have $G(\omega_0) \to 1$. We note that reducing $\omega_0$ at fixed $\lambda / J =0.2$ substantially suppresses the absolute value of the string correlator because the effective coupling to the bath $\propto \lambda^2/\omega_0$ increases, but also because $\omega_0 \to 0$ represents a mean-field limit in which long-range AFM order can occur that suppresses the string-order parameter.

\begin{figure}
  \centering
\includegraphics[width=0.85\linewidth]{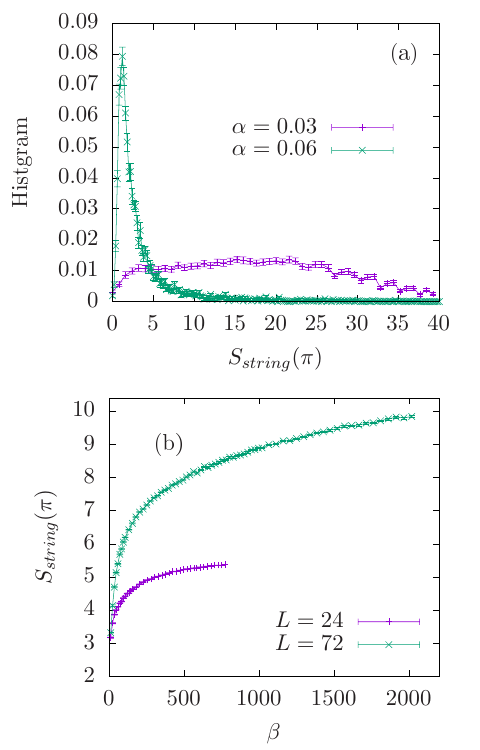}
\caption{\label{fig:Hist}%
(a) Histograms of $C_\mathrm{string}(q=\pi)$ calculated from individual QMC configurations. Here $L=96$ and $\beta J=192$. We choose $\alpha=0.03$ within the Haldane phase and $\alpha=0.06$ in the AFM phase.
(b) Inverse-temperature dependence of $S_\mathrm{string}(q=\pi)$ for $L=24$ and $L=72$. We choose  a coupling $\alpha=0.04348$ close to the bulk critical point.  
  }
\end{figure}

Having tested our QMC estimator for a single boson frequency $\omega_0$, we now want to return to our ohmic bath with a continuous spectrum. The sampling of our bosonic spectral function $J(\omega)$, as described in Ref.~\cite{PhysRevB.105.165129}, creates a distribution of $\omega_{k}$. Because we need to calculate the product of $G(\omega_{k})$ in Eq.~\eqref{eq:QMC_parity}, already a single $\beta \omega_{k} \ll 1$ can substantially suppress the local contribution to the string correlation function.
To convince ourselves that this does not pose a problem, we first take a look at the distribution of measurements for the string order parameter $C_\mathrm{string}(q=\pi)$. Figure \ref{fig:Hist}(a) shows histograms of single QMC measurements of $C_\mathrm{string}(q=\pi)$ in the Haldane phase and in the AFM phase at $L=96$ and $\beta J =192$. We observe a broad distribution of measurements in the Haldane phase which will result in a rather large mean value, whereas in the AFM phase measurements are peaked at small values because string order is suppressed. The temperature dependence of $C_\mathrm{string}(q=\pi)$ is illustrated in Fig.~\ref{fig:Hist}(b) close to the bulk critical point and for two system sizes $L=24$ and $72$. As expected, temperature convergence is significantly slower for the continuous spectrum than for the single bosonic frequency because there are always some vertices with low frequencies. 
However, we always obtain a finite estimate that evolves smoothly with temperature. Although doubling $\beta$ increases the chance of finding a vertex at a low frequency $\omega_{k}$, the weight $G(\omega_{k})$ is determined by the product $\beta \omega_{k}$ so that the two factors outweight each other.
Moreover, the slow temperature convergence is not a problem for our analysis of quantum criticality, because close to the critical point it is only important to scale $\beta \propto L^z$ appropriately. Our precise estimate of the critical coupling via the string correlation ratio confirms the accuracy of our estimator.

\section{Epsilon expansion for ordinary boundary criticality}

For the dissipative $\phi^4$ theory discussed in the main part of our paper, the $\epsilon$ expansion was used in Ref.~\cite{Sachdev_2004_NFL} to calculate the critical exponents of the bulk quantum phase transition driven by the coupling to an ohmic bath. Here, we extend this calculation to derive the boundary critical exponents. 
Our derivation applies to the \textit{trivial} case where no additional Berry phase terms at the boundary are taken into account. This scenario corresponds to the transition from the trivial dimerized phase to the AFM state, but also to the transition in the dissipative quantum rotor model, for which we will discuss the boundary criticality in the next section.

Consider the action $\mathcal{S} = \mathcal{S}_0 + \mathcal{S}_\mathrm{diss} + \mathcal{S}_\mathrm{surf}$
defined on a semi-infinite geometry. To this end,
we split our spatial coordinates
$\mathbf{x}=(\mathbf{x}_\parallel, x_\perp)$ into a $\dimred$-dimensional component $\mathbf{x}_\parallel \in \mathds{R}^\dimred$ parallel to the boundary and a one-dimensional part $x_\perp \in [0,\infty)$ perpendicular to it (here we have defined $\dimred = d-1$). Then, our $\phi^4$ theory with $N$-component vectors $\boldsymbol{\phi}$ reads
\begin{align}
\nonumber
\mathcal{S}_0
=
& \int
d^\dimred \mathbf{x}_\parallel \,
d x_\perp \, d\tau \,
%\int\limits_{-\infty}^\infty
%d^d x_\parallel \int\limits_0^\infty
%d x_\perp \int\limits_0^\infty d\tau \,
\bigg\{ \frac{1}{2}\left[\nabla \boldsymbol{\phi}  ( \mathbf{x}_\parallel, x_\perp, \tau ) \right]^2
\\
& + \frac{r}{2} \boldsymbol{\phi}(\mathbf{x}_\parallel, x_\perp, \tau)^2 
+ \frac{g}{4!} \left[ \boldsymbol{\phi}(\mathbf{x}_\parallel,x_\perp,\tau)^2 \right]^2
\bigg\} \, ,
\end{align}
whereas the effects of ohmic dissipation are captured by the local but frequency-dependent contribution 
\begin{align}
\mathcal{S}_\mathrm{diss}
=
\int d^\dimred \mathbf{x}_\parallel \, d x_\perp \, 
  d \omega \, |\omega| \left| \boldsymbol{\phi}( \mathbf{x}_\parallel, x_\perp, \omega ) \right|^2
  \, .
\end{align}
Moreover, we include the surface contribution
\begin{align}
\mathcal{S}_\mathrm{surf}
=
\frac{c}{2} \int d^\dimred {\mathbf{x}_{\parallel}} \,  d\tau \, \boldsymbol{\phi} ( \mathbf{x}_\parallel, 0, \tau )^2
\, ,
\end{align}
which is tied to the boundary at $x_\perp = 0$.

To study the boundary criticality of the ordinary phase transition which occurs at $r=0$, it is sufficient to fix $c=\infty$ so that the amplitude of $\phi$ along the boundary is zero.
For the ohmic bath, the upper critical dimension is $\dimred=1$ (\ie, the total spatial dimension is 2), and we perform an $\epsilon = 1 - \dimred$ expansion about this dimension. Our derivation of the ordinary boundary criticality closely follows Ref.~\cite{Diehl:1981jgg}, where the same calculation has been performed for the O($N$) model without a dissipative term. Therefore, we only sketch the main steps in the following.

Our calculation for the boundary criticality of the dissipative case builds up on the $\epsilon$ expansion for the bulk critical point performed in Ref.~\cite{Sachdev_2004_NFL}. We will not repeat these steps here, but we will make use of some of their results. In particular, for the renormalization of the coupling $g$ we introduce the normalized constant $u$ by rewriting  
\begin{equation}
   g = Z_u s_\dimred^{-1} u  \, . 
\end{equation} 
Here we defined $s_\dimred \equiv 2^{-\dimred+1} \pi^{-\dimred/2} $ and introduced the coupling-constant renormalization function $Z_u$ which is  
\begin{equation}
   Z_u  = 1 + \frac{ n+8 }{ 3 \epsilon }u + O( u^2 ) \, .  
\end{equation}

For the ordinary transition, the boundary suppresses the order parameter fluctuations at the surface. Therefore, the leading boundary  operator is its normal derivative  $\partial_{\perp} \phi$, which determines the surface critical behavior. We define
\begin{equation}
  \phi^{\perp}( \bm x_\parallel, \tau ) \equiv \partial_{\perp} \phi( \bm x_\parallel, 0^{+}, \tau ) 
\end{equation} 
as the derivative of the order parameter along the direction that is perpendicular to the surface.

In the presence of a boundary, translational invariance is broken in the perpendicular direction. Therefore, correlation functions involving boundary operators may contain additional ultraviolet divergences that cannot be absorbed by the bulk renormalization constants $Z_\phi$. A new renormalization factor $Z_1$ 
is thus required for the surface operator, \ie, 
 \begin{equation}
  \phi^{ \perp } = ( Z_\phi Z_1 )^{1/2} \phi^{\perp}_R    
 \end{equation}
where $ \phi^{\perp}_R $ is the renormalized surface operator and $Z_\phi$ is the field renormalization constant inside the bulk. 

The divergence of the correlation function between a surface operator and a bulk operator, 
\begin{equation}
  G^{ \text{surf} }  \equiv \langle \phi (\mathbf{x}_\parallel, z, \tau )  \phi^\perp (\mathbf{x}_{ \parallel }, 0, \tau)  \rangle  \, ,
\end{equation}
can be absorbed by the renormalization function $Z_1$. 
We define the corresponding anomalous dimension at the boundary, \ie,
\begin{equation}
  \eta_1 ( u ) \equiv 
  \beta ( u) \, 
  \partial_{ u } |_0 \ln Z_1  \, , 
\end{equation}
where $\beta(u) = - \epsilon u + \frac{N+8}{3} u^2 + \mathcal{O} (u^3) $ is the bulk beta function.
The boundary critical exponent satisfies the scaling relation 
\begin{equation}
\label{eq:eta_rel}
   \eta_{ \parallel } = 2 + \eta + \eta_1^{*} \, .
\end{equation}
We assume that the two-loop simple pole contribution to $Z_1$ vanishes, as for the ordinary boundary transition of the non-dissipative $\phi^4$ theory \cite{Diehl:1981jgg}. Under this assumption, the $\mathcal{O}(\epsilon^2)$ correction to the boundary critical exponent arises solely from the bulk contribution.  In the following, we outline the one-loop integration to extract $Z_1$. 
The calculation is carried out in momentum space. We label $\bm{p}$ ($k$) as the momentum along the parallel (perpendicular) direction. Naturally, momentum conservation is preserved  for  
$\bm{p}$. 
The surface operator is then  denoted as 
\begin{equation}
   \phi^{ \perp } (\bm p, \omega) \equiv \int d k \, k \,  \phi ( \bm{p}, k, \omega)  
\end{equation}
such that the zeroth-order correlation function reads 
 \begin{align}
 \nonumber
  G^{ \text{ surf } }_0 (\bm{p}, k, \omega) &= \langle \phi ( \bm p, k, \omega) \, \phi^{ \perp } ( -\bm p, \omega) \rangle_0  \\ 
   &= \frac{k}{ |\bm{p}|^2 + |\omega| + k^2 } 
\end{align}

The one-loop correction is denoted as 
\begin{align} 
   G^{ \text{surf} }_1(\bm{p}, k, \omega)  
 = & \frac{(N+2) g}{6 \, (2\pi)^{\dimred+3} }  \int d^\dimred \widetilde{\bm{p}}  \int d \widetilde{\omega}    \int d \widetilde{k}  
  \int d k_1   \nonumber\\ 
 \times & \frac{1}{ |\bm p| ^2 + |\omega| + k^2 }   \frac{ k_1 }{  |\bm p|^2 + |\omega| + k_1^2 }  \nonumber\\   \times &
  \frac{1}{ |\widetilde{ \bm p }| ^2 +  |\widetilde{\omega}| + \widetilde{k}^2 } 
   \pi
  \delta( k + k_1 + 2 \widetilde{ k }  )  
\end{align}     
where $ \bm{p}$ ($k$) denotes the parallel (perpendicular) momentum of the external line, and $\omega$ is the corresponding frequency.  
Note that only $\bm{p}$ and $\omega$ respect momentum (frequency) conservation as in usual systems with translational invariance. 
 The corresponding parallel momentum and frequency of the loop line, marked as $\widetilde{\bm p}$ and $\widetilde \omega$,  are free variables 
 to be integrated out.  
 Note that translational invariance does not hold in general for momentum $k$, due to the boundary condition.  
 The ordinary boundary condition $c \to \infty$ leads to the  vertex constraint  
$ \pi  \delta( k + k_1 + 2 \widetilde{ k } ) $, where $\widetilde{k}$ is the momentum of the loop line.

We first perform integration with respect to $ \widetilde{\bm p } $ and $\widetilde{\omega}$, \ie, 
\begin{align}
\nonumber
   & \frac{1}{ (2\pi)^{\dimred+1} } \int d^\dimred  \widetilde{ \bm p}  \int d \widetilde{ \omega } \, \frac{1}{  |\widetilde{ \bm p}|^2 + |\widetilde \omega| + [( k_1 + k )/2]^2 }   \\  
 & \quad=  \frac{(-1)}{ 2^{ 2\dimred-1 } \pi^{ 1+ \dimred/2 } } \frac{1}{\dimred}   \Gamma( 1-\dimred /2 )  [ ( k_1 + k ) ]^{\dimred}  
\end{align}   
To integrate over $ k_1 $, we consider Feynman parametrization and introduce $\epsilon = 1-\dimred$ such that
\begin{align}
   & \frac{1}{| \bm p|^2 + |\omega| + k_1^2 } \frac{1}{  (  k_1 +   k )^{ -(1 - \epsilon) } }   \nonumber\\ 
 & \quad = \frac{ \Gamma( \frac{1+\epsilon}{2} ) }{  \Gamma (  -\frac{1-\epsilon}{2}  ) }   \int_0^1 d \alpha \, \alpha^{ (\epsilon-3)/2 }  
  \big[ \alpha ( k_1 + k )^2 \nonumber\\ 
 & \qquad\qquad+ (1 -\alpha) ( |\bm p|^2 + |\omega| + k_1^2) \big]^{ -(1+\epsilon)/2 }  \, .
\end{align}    
As a next step, we need to evaluate the integral
\begin{align} 
I =  &  \frac{1}{2\pi} \int_0^1 d \alpha \, \alpha^{ ( \epsilon-3 )/2 } \int d k_1   k_1  
 [ ( k_1 + A )^2 + M^2 ]^{ -( \epsilon+1 )/2 }  \nonumber\\ 
  = &   \int_0^1 d\alpha \,  \alpha^{ ( \epsilon-3 )/2 } \frac{ (-A) }{ \sqrt{ 4  \pi} }  
  \frac{ \Gamma( \epsilon/2 ) }{ \Gamma ( \frac{ \epsilon + 1}{2} ) } ( M^2 )^{ - \frac{\epsilon}{2} }  
\end{align} 
where we have defined 
\begin{align}
A &\equiv \alpha  k \, , \nonumber\\
M^2 &\equiv (1- \alpha) ( |\bm p|^2 + \omega + \alpha k^2) \, .
\end{align}
Integrating over $\alpha$   
and performing the expansion in $\epsilon$, we obtain
\begin{equation}
\begin{aligned} 
  I = & -  \frac{k}{ 2 \sqrt{\pi} }   \left[ \frac{2}{\epsilon} - C_E + O(\epsilon) \right]  \frac{1}{ \sqrt{\pi} }   
  \int_0^1  d\alpha \, \alpha^{ -\frac{1}{2} }  \\ 
   &\times \left[ 1+\frac{\epsilon}{2} \ln \alpha - \frac{\epsilon}{2} \ln (1 - \alpha ) 
  - \frac{\epsilon}{2} \ln ( |\bm p|^2 + \omega + \alpha k^2 ) \right]   \\ 
  = & -\frac{ 2k }{ \pi } \left( \frac{1}{\epsilon} -F \right) + \mathcal{O}( \epsilon ) 
\end{aligned} 
\end{equation}
where we have defined ($C_E$ is the Euler constant)
\begin{equation}
   F \equiv  \frac{|\bm p |}{k} \arctan\left(\frac{k}{|\bm p|}\right)  +
    \frac{1}{2} \ln( |\bm p|^2 + \omega + k^2 ) +
    \frac{ C_E }{2}  - 1 \, .
\end{equation}
Altogether,  the one-loop correction becomes 
\begin{equation}
   C^{ \text{surf} }_1 = \frac{k}{ |\bm{p}|^2 + |\omega| + k^2 }  \frac{g s_\dimred }{ 2 } \frac{ (n+2) }{3} \left( { \frac{1}{ \epsilon } - \frac{F}{2}} \right) + \mathcal{O}( \epsilon ) \, .
\end{equation}

All in all, we obtain the expansion form of the boundary renormalization function,  
\begin{equation} 
  Z_1 = 1 + \frac{ n+2 }{ 3 \epsilon } u + \mathcal{O}( u^2, \epsilon^2 ) \, ,
\end{equation}
such that 
\begin{equation}
  \eta_1 (u) = - \frac{n+2}{3}  u + \mathcal{O}(u^3) \, .
\end{equation} 
At this point, we need to use that the bulk fixed point up to quadratic order in $\epsilon$ reads
\begin{equation}
\begin{aligned}
 u^* = \frac{3}{N+8}\epsilon + \frac{3 [ ( 14 -\pi^2/3 )N + 76 - 8\pi^2 /3  ] }{(N+8)^2} \epsilon^2   
\end{aligned}    
\end{equation}
such that we obtain
\begin{align} 
 \eta_1(u^*) = & -\frac{ N+2 }{ N +8 } \epsilon \\ 
  - & \frac{ (N+2) [ (14 - \pi^2/3 )N + 76  - 8 \pi^2 / 3 ] }{ (N+8)^3 }  \epsilon^2  \, .
  \nonumber
\end{align}
Plugging $\eta_1(u^*)$ as well as the bulk anomalous dimension
  \begin{equation} 
   \eta = \frac{ ( 12 - \pi^2 ) (N+2) }{ 4 (N+8)^2 }  \epsilon^2 + \mathcal{O}(\epsilon^3)
\end{equation}
into Eq.~\eqref{eq:eta_rel}, we finally obtain the boundary anomalous dimension
\begin{align}
\label{eq:etapara}
  \eta_\parallel = & 2 - \frac{N+2}{N+8} \epsilon \\ - & \frac{(N+2) [ (44-4\pi^2/3)N + 208-8\pi^2/3 ] }{4(N+8)^3}  \epsilon^2  \, .
  \nonumber
\end{align}
We want to clarify again that we have only explicitly calculated the one-loop correction of the boundary renormalizaton. 
The two-loop boundary diagrams vanish for the classical O($N$) model \cite{Diehl:1981jgg} and it remains for future studies to check if this is also the case for the dissipative O($N$) model. In any case, the $\epsilon^2$ contribution of the bulk renormalization is an additive correction which we should take into account and which leads to good agreement of Eq.~\eqref{eq:etapara} with numerics.

\section{Estimation of critical exponents}

In the following, we present the details of how we estimated our critical exponents. Moreover, we discuss additional results for the boundary criticality.

\subsection{Bulk criticality}

\subsubsection{Critical coupling and correlation-length exponent}

We first estimate the bulk critical coupling using the correlation ratio $R$ 
which becomes renormalization-group invariant at $\alphac$. In the absence of scaling corrections and at fixed $\beta / L^z$, it fulfills the scaling ansatz
\begin{align}
\label{eq:scaling}
R(\alpha,L) = f(L^{1/\nu}(\alpha-\alphac)) \, .
\end{align}
If we calculate $R(\alpha)$ for different system sizes $L$, Eq.~\eqref{eq:scaling} implies that all curves should cross at the critical coupling $\alphac$.
However, in the presence of scaling corrections, the crossings between data pairs $(L-\Delta L, L)$ exhibit a finite-size drift $\alphac(L) = \alphac + A \, L^{-e}$ where $e = 1/\nu + \omega$ contains the exponent $\omega$ of the leading scaling correction and $A$ is a nonuniversal constant. 
In practise, we fit data pairs $(L-\Delta L, L)$ to the Taylor-expanded scaling form %\cite{Toldin14}
\begin{align}
\label{eq:Rfit}
    R (\alpha, L) = R^* + \sum_{n=1}^{ n_{ \text{max} } } a_n \left( \alpha - \alphac \right)^n L^{n / \nu }  
\end{align}
with $n_\mathrm{max}=2$, where $R^\ast$, $a_n$, $\alphac$, and $\nu$ are treated as fitting constants. Fitting pairs of system sizes to Eq.~\eqref{eq:Rfit} has the advantage that we get access not only to $\alphac(L)$ but also to the correlation-length exponent which can be extrapolated as $\nu(L)=\nu + A \, L^{-\omega_\nu}$.
Note that we fix $z=2$ for all calculations of critical exponents but that we have checked its validity in our main paper via the local susceptibility.

\begin{figure}[t]
  \centering
\includegraphics[width=0.5\textwidth]{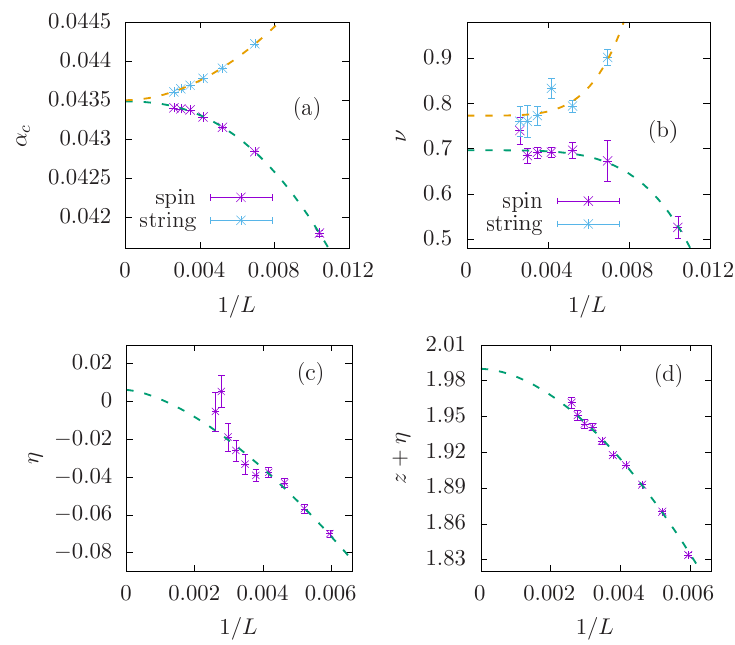}    \caption{\label{figSM:bulk_Haldane}%
Finite-size analysis at the bulk quantum phase transition between Haldane and AFM states at $\delta=0$, as described in the main text. Using power-law fits, we extrapolate (a) $\alphac$ and (b) $\nu$, both estimated from $R_{ \text{spin} }$ and $R_{\text{string}}$, as well as (c) $\eta$ and (d) $z+\eta$ obtained from $\chi_{\text{spin}}(q=\pi)$ and $C_{\text{spin}}(r=L/4)$, respectively.} 
  \end{figure}

For the Haldane chain, the main part of our paper already included a plot of the correlation ratios calculated from the spin and string correlation functions. A finite-size extrapolation of their pseudocritical couplings $\alphac(L)$ is shown in Fig.~\ref{figSM:bulk_Haldane}(a). We find that both of them extrapolate to the same critical coupling, indicating that there is a direct transition from the Haldane phase to the AFM state. A finite-size extrapolation of the correlation-length exponent $\nu(L)$ is depicted in Fig.~\ref{figSM:bulk_Haldane}(b). We extrapolate $\nu=0.697(8)$ from the spin correlations and $\nu=0.77(3)$ from the string correlations. We note that the string correlations suffer from larger error bars but its estimate is still in agreement with the one from the spin correlations.

\begin{figure}[t]
  \centering
\includegraphics[width=0.5\textwidth]{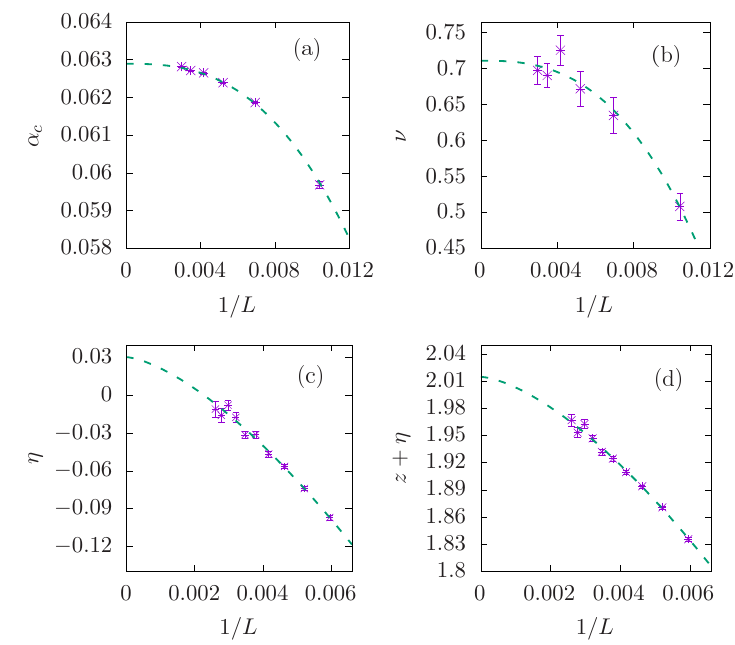}    \caption{\label{figSM:bulk_dimer}%
Same as Fig.~\ref{figSM:bulk_Haldane}, but for the bulk quantum phase transition between the trivial and the AFM state at $\delta=0.4$. 
}
  \end{figure}

We have repeated the same analysis for the transition from the trivial phase to the AFM state at $\delta=0.4$. Our extrapolation of $\alphac(L)$ and $\nu(L)$ is shown in Figs.~\ref{figSM:bulk_dimer}(a) and \ref{figSM:bulk_dimer}(b). Our estimate $\nu=0.71(2)$ is in good agreement with our previous results.

\subsubsection{Anomalous dimension and scaling dimension}

Right at the bulk critical point $\alpha = \alphac$, we can estimate a variety of critical exponents using the same technique: Consider an observable $O \propto L^\mu \left( 1+g \, L^{-\omega_\mu} \right)$ that scales with a dimension $\mu$. 
Using a pair of system sizes $(L-\Delta L, L)$, we can calculate a finite-size estimate
\begin{align}
\label{eq:extract_exp}
\mu(L) = \frac{1}{\ln r_L} \ln \frac{O(r_L L)}{O(L)}
\end{align}
with $r_L=(L-\Delta L)/L$ and subsequently extrapolate $\mu(L)$ using a power-law fit. For example, we use this technique to extract the bulk anomalous dimension $\eta$ from $\chi(q=\pi) \propto L^{2-\eta}\left(1+g \, L^{-\omega_\eta}\right)$.
In the same way, we can use the real-space correlations
$C(r) \propto  r^{ -2 \Delta } f(r/L)$ at fixed $r/L$ to estimate the scaling dimension $\Delta =(d+z-2+\eta)/2$ or, equivalently, $z+\eta$. For both cases, we fix $\Delta L = 120$.

Finite-size extrapolation of $\eta$ and $z+\eta$ is performed in Figs.~\ref{figSM:bulk_Haldane}(c) and \ref{figSM:bulk_Haldane}(d) for the transition out of the Haldane phase as well as in Figs.~\ref{figSM:bulk_dimer}(c) and \ref{figSM:bulk_dimer}(d) for the transition out of the trivial phase. Again, our estimates for the critical exponents agree well with each other.
All bulk critical exponents are tabulated in the main part of our paper.

\subsection{Boundary criticality}

\subsubsection{Correlation functions and scaling dimensions}
\label{sec:Cfunctions}

At the bulk critical point, correlation functions
\begin{align}
C(i,j;\tau) = \langle S_i^x(\tau) S_j^x(0)  \rangle
\end{align}
experience distinct asymptotic behavior depending on whether zero, one, or two of the spin operators are placed in the vicinity of the boundary. We therefore distinguish three cases: (i) If both $i$ and $j$ are far away from the boundary, we only probe the bulk properties. Then,
\begin{align}
C_\mathrm{bulk}(r,\tau)
    =
    C(r_0+r,r_0;\tau)
    \propto
    r^{-2 \Db} f( r^z / \tau ) 
\end{align}
with $f$ being a universal function, 
is determined by the bulk scaling dimension $\Db = (d+z-2+\eta)/2$.
(ii) If both $i=j=1$ are located at the boundary of our system, we can probe the edge correlation function
\begin{align}
\label{eq:Cbound}
C_\mathrm{edge}(\tau)
    =
    C(1,1;\tau)
    \propto
    \tau^{-2\De / z}
\end{align}
along imaginary time. Its power-law decay is determined by the edge scaling dimension $\De = (d+z-2+\eta_\parallel)/2$.
(iii) If only one of the operators is located at the boundary, we can measure the mixed correlation function
\begin{align}
\label{eq:Cbound-bulk}
C_\mathrm{edge-bulk}(r) = C(1+r,1;0)
\propto
r^{-(\Db + \De)}
\end{align}
at equal times, which is given by the sum of bulk and boundary scaling dimensions. Consequently, the anomalous dimension of the edge-bulk correlations, $\eta_\perp = (\eta + \eta_\parallel)/2$, is not independent.

\subsubsection{Transition out of the Haldane phase}

\begin{figure*}[t]
  \centering
    \includegraphics[width=0.45\textwidth]{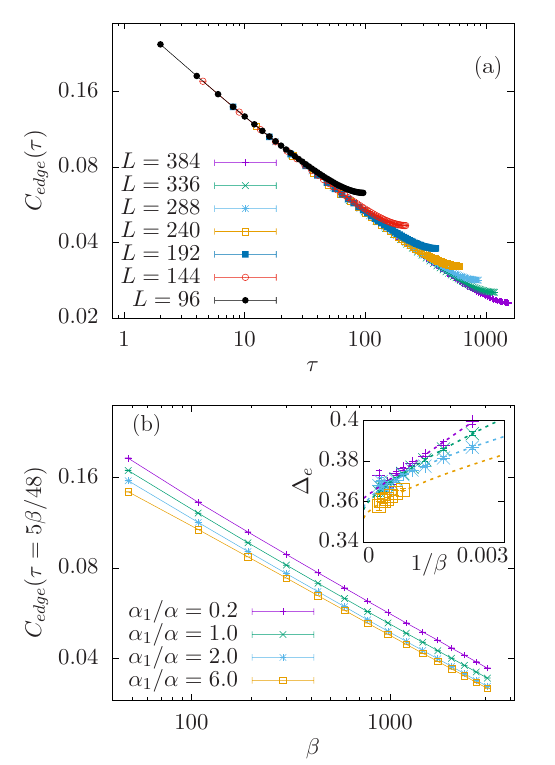}
    \includegraphics[width=0.45\textwidth]{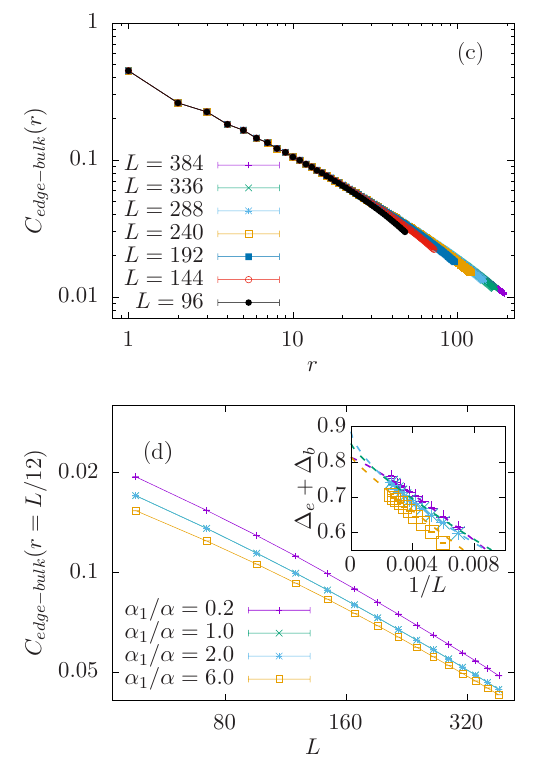}
    \caption{\label{figSM:boundary_Haldane}%
Boundary criticality at the quantum phase transition between Haldane and AFM states at $\delta=0$, as extracted from the spin-spin correlation functions using $\beta J = L^2/48$.
Edge correlations: (a) $C_\mathrm{edge}(\tau)$ as a function of imaginary time $\tau$ for various system sizes $L$. (b) $C_\mathrm{edge}(\tau = 5\beta/48)$ as a function of $\beta$ for different boundary couplings $\alpha_1/\alpha$. We use data pairs $(L-\Delta L, L)$ with $\Delta L= 120$ to extract a finite-size estimate of scaling dimension $\Delta_\mathrm{e}$ as depicted and extrapolated in the inset.
Edge-to-bulk correlations: (c) $C_\mathrm{edge-bulk}(r)$ as a function of distance $r$ for various $L$. (d) $C_\mathrm{edge-bulk}(r=L/12)$ as a function of $L$ for different boundary couplings $\alpha_1/\alpha$. The inset shows a finite-size extrapolation of the scaling dimension $\Delta_\mathrm{b} + \Delta_\mathrm{e}$.
     }
  \end{figure*}

We first consider the transition from the Haldane phase to the AFM state at $\delta=0$
and extract the boundary scaling dimension from the spin-spin correlation functions defined in Eqs~\eqref{eq:Cbound} and \eqref{eq:Cbound-bulk}. Note that for all simulations we keep $\beta J = L^2/48$ fixed.

Figure~\ref{figSM:boundary_Haldane}(a) shows $C_\mathrm{edge}(\tau)$ as a function of imaginary time for various system sizes $L$. All data sets fall on top of each other (apart from boundary effects that become visible near $\tau=\beta/2$) and exhibit a power-law decay for which subleading corrections appear to be small. To perform a quantitative finite-size-scaling analysis, we plot $C_\mathrm{edge}(\tau)$ at fixed $\tau=5\beta/48$ as a function of inverse temperature $\beta$ in Fig.~\ref{figSM:boundary_Haldane}(b). Because $\beta J = L^2/48$ is fixed, this is equivalent to a scaling analysis in system size.
We use the procedure defined by Eq.~\eqref{eq:extract_exp} to extract a size dependent scaling dimension $\De(L)$ from data pairs $(L-\Delta L, L)$ with $\Delta L = 120$. The latter is shown in the inset of Fig.~\ref{figSM:boundary_Haldane}(b) as a function of $1/\beta$ and extrapolated towards $\beta\to \infty$ using a power-law fit. We repeat this analysis for different boundary couplings $\alpha_1 / \alpha \in\{0.2, 1, 2, 6\}$ which all lead to consistent estimates of $\De$, as collected in Table~\ref{tab:scalingdim}.
The scaling dimensions can be translated into the boundary anomalous dimensions $\eta_\parallel$ via the relation $2 \De = (d+z-2+\eta_\parallel)$.

We also want to extract the boundary scaling dimension from the spin-spin correlation function between edge and bulk, $C_\mathrm{edge-bulk}(r)$, which is shown in Fig.~\ref{figSM:boundary_Haldane}(c) as a function of distance $r$ and for various system sizes. Already at this stage we can observe that $C_\mathrm{edge-bulk}(r)$ exhibits a stronger crossover as function of distance than $C_\mathrm{edge}(\tau)$. To perform a quantitative finite-size scaling, we plot $C_\mathrm{edge-bulk}(r=L/12)$ as a function of $L$ in Fig.~\ref{figSM:boundary_Haldane}(d)
and extract a size-dependent scaling dimension as before. Its finite-size extrapolation is performed in the inset of Fig.~\ref{figSM:boundary_Haldane}(d) for different boundary couplings $\alpha_1/\alpha$ and our extrapolated estimates for $\Db + \De$ are collected in Table \ref{tab:scalingdim}. Because finite-size drifts are larger in this case, our estimates of the scaling dimension have larger error bars. Overall, our estimate of $\Db + \De$ is in good agreement with the independent estimates of $\Db$ and $\De$.

\begin{table}%
\caption{\label{tab:scalingdim}%
Comparison of the scaling dimensions determined from the edge, edge-bulk, and bulk correlation functions for the Haldane chain, the dimerized chain, and the quantum rotor model with ohmic dissipation. We also vary the boundary coupling $\alpha_1/\alpha$ and include the predictions of the $\epsilon$ expansion.
}
\begin{ruledtabular}
\begin{tabular}{c|c|c|c|c}
 & $\alpha_1 / \alpha$ & 
 $\De$ &
 $\Db + \De$ &
 $\Db$ \\
\colrule
        & 0.2 & 0.362(5) & 0.81(2) &   \\
Haldane & 1.0 & 0.356(5) & 0.84(2) & 0.49(1)  \\
chain   & 2.0 & 0.357(6) & 0.88(2) &   \\
        & 6.0 & 0.35(3)  & 0.82(2) &   \\
\colrule
          & 1.0 & 1.14(1) & 1.72(4) &   \\
dimerized & 2.0 & 1.11(2) & 1.75(6) & 0.51(1)  \\
chain     & 4.0 & 1.15(4) & 2.1(3) &   \\
          & 6.0 & 1.26(7) & 2.6(6) &   \\
\colrule
O(3) model & 1.0 & 1.16(2) & 1.55(3) & 0.52(2)   \\ 
\colrule
$\epsilon$ expansion & -- & 1.144 & 1.644 &  0.5  \\

\end{tabular}
\end{ruledtabular}
\end{table}

\subsubsection{Transition out of the dimerized phase}

\begin{figure*}[t]
  \centering
  \includegraphics[width=0.45\textwidth]{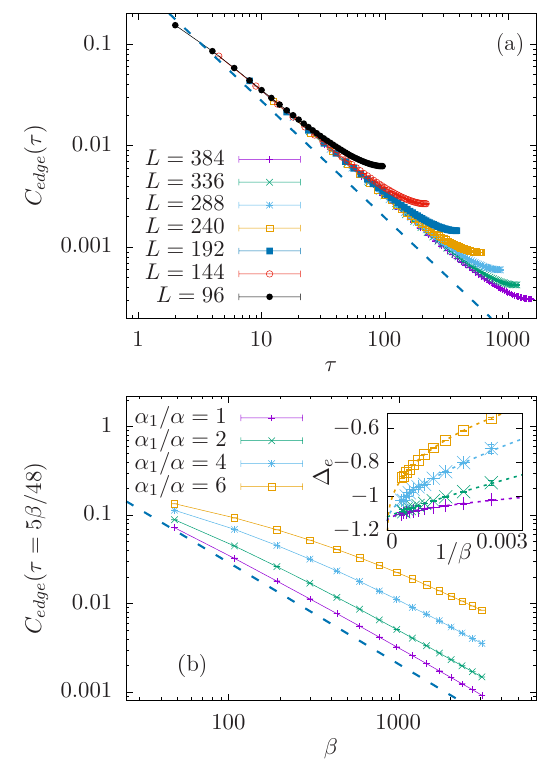}
  \includegraphics[width=0.45\textwidth]{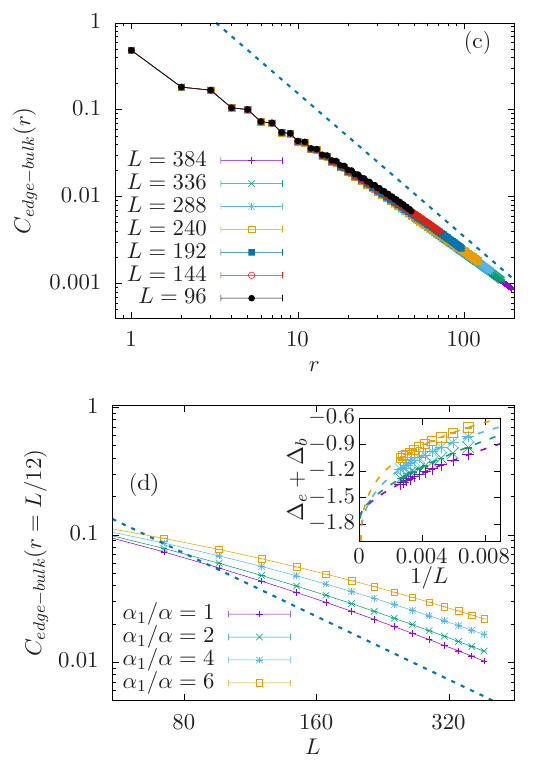}
    \caption{\label{figSM:boundary_dimer}%
Boundary criticality at the quantum phase transition between dimerized and AFM states at $\delta=0.4$, as extracted from the spin-spin correlation functions using $\beta J = L^2/48$.
As described in the caption of Fig.~\ref{figSM:boundary_Haldane},
we display the edge correlations in (a), (b) and the edge-to-bulk correlations in (c), (d), including an extrapolation of the corresponding scaling dimensions in the insets. The dashed lines in (a)--(d) display power-law decays with the scaling dimensions extracted from the $\epsilon$ expansion.
  }
  \end{figure*}

In Fig.~\ref{figSM:boundary_dimer} we repeat the same analysis for the quantum phase transition from the topologically-trivial dimerized phase to the AFM state at fixed $\delta=0.4$. Again, our estimates for the scaling dimensions are collected in Table~\ref{tab:scalingdim}, but now they can be compared to the predictions of the $\epsilon$ expansion. Our estimates for $\De$ from $C_\mathrm{edge}(\tau)$ are in excellent agreement with the $\epsilon$ expansion. However, we can see from Fig.~\ref{figSM:boundary_dimer}(b) that an increase of the boundary coupling $\alpha_1/\alpha$ significantly increases the scaling corrections. Crossover effects become even more dramatic when analyzing $C_\mathrm{edge-bulk}(r)$ in Figs.~\ref{figSM:boundary_dimer}(c) and \ref{figSM:boundary_dimer}(d) for which our numerical data show substantial deviations from the dashed lines that indicate the prediction of the $\epsilon$ expansion. While finite-size extrapolations are still in agreement with the analytical predictions, error bars are significantly larger, especially for strong $\alpha_1 / \alpha$. It is therefore better to extract the independent boundary scaling dimension $\De$ from $C_\mathrm{edge}(\tau)$.

\subsubsection{Dissipative quantum rotor model}

\begin{figure*}[t]
  \centering
  \includegraphics[width=0.45\textwidth]{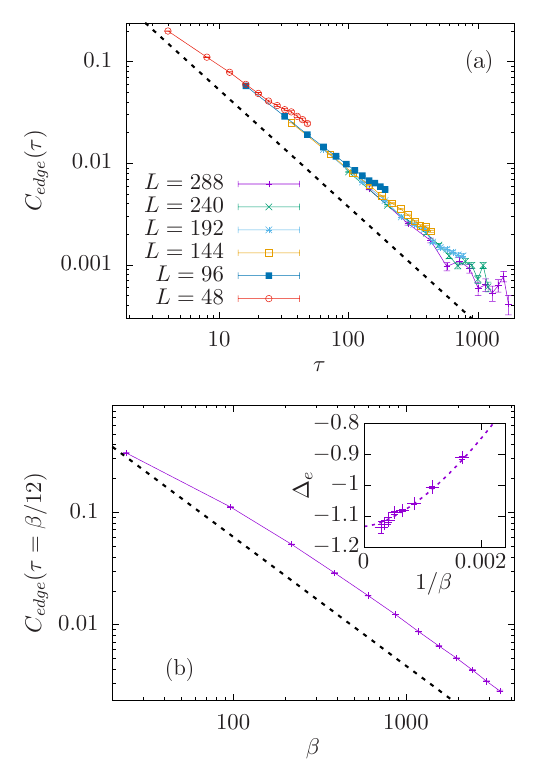}
  \includegraphics[width=0.45\textwidth]{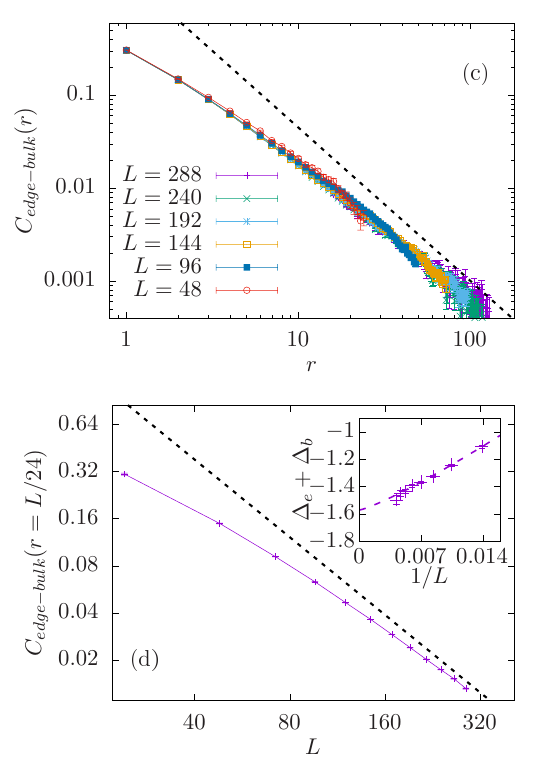}
    \caption{\label{figSM:rotor}
Boundary criticality in the dissipative O(3) quantum rotor model.
Edge correlations:
(a) $C_\mathrm{edge}(\tau)$ as a function of imaginary time $\tau$ for various system sizes $L$. (b) $C_\mathrm{edge}(\tau = \beta/12)$ as a function of $\beta$. We use data pairs $(L-\Delta L, L)$ with $\Delta L= 120$ to extract a finite-size estimate of scaling dimension $\Delta_\mathrm{e}$ as depicted and extrapolated in the inset.
Edge-to-bulk correlations:
(c) $C_\mathrm{edge-bulk}(r)$ as a function of distance $r$ for various $L$. (d) $C_\mathrm{edge-bulk}(r=L/24)$ as a function of $L$. The inset shows a finite-size extrapolation of the scaling dimension $\Delta_\mathrm{b} + \Delta_\mathrm{e}$. 
The dashed lines in (a)--(d) display power-law decays with the scaling dimensions extracted from the $\epsilon$ expansion.
  }
  \end{figure*}

The transition from the dimerized to the AFM phase is described by the O(3)-symmetric nonlinear sigma model without topological term but with ohmic dissipation. 
Its action can be discretized (here $\Delta \tau = \beta / N_\tau$),
 \begin{align}
 \nonumber
 \mathcal{S} =& -\sum_{i = 1}^L \sum_{ \tau = 1 }^{N_\tau}  \left[ K_x \, \mathbf{n}_{i, \tau} \cdot \mathbf{n}_{i+1, \tau} + K_\tau \, \mathbf{n}_{i, \tau} \cdot \mathbf{n}_{i, \tau+1} \right] \\
       & -   \alpha  \sum_{i = 1}^L  \sum_{ \tau < \tau' }  \bigg( \frac{\pi}{ N_{\tau} } \bigg)^2
      \frac{\mathbf{n}_{i, \tau} \cdot \mathbf{n}_{i, \tau'}}{\sin^2 \!{\big(\frac{\pi}{N_\tau}|\tau-\tau'|\big)}} \, ,
    \label{eq:Srotor}
 \end{align}
 where $\mathbf{n}_{i, \tau}$ is a $3$-component vector at site $i\in\{1,\dots, L\}$ and imaginary-time slice $\tau \in\{1, \dots, N_\tau\}$ that satisfies $ |\mathbf{n}_{i, \tau}|^2 = 1 $. This action also corresponds to a dissipative quantum rotor model which,
 due to the quantum-to-classical mapping, can be simulated directly using classical Monte Carlo methods with cluster updates \cite{LuijtenBloethe95}.
 Previously, only the bulk transition of the Ising and O(2)-symmetric cases has been studied in this way \cite{Werner_2005_ising, Werner_2005}.
 
 Here we are interested in the boundary response of the O(3)-symmetric model.
 For our numerical simulations, we set
$K_x=0.5$, $K_\tau=0.3517$, and $N_\tau = L^2/12$, for which we have estimated the bulk critical coupling $\alphac = 0.7158(1)$ in the same way as described for the quantum spin chain.
For our analysis, we calculate the
space-time correlation functions
\begin{align}
\label{eq:Crotor}
C(i,j;\tau)
    =
    \frac{1}{3} \langle \mathbf{n}_{i,\tau} \cdot \mathbf{n}_{j,0} \rangle
    \, .
\end{align} 
From Eq.~\eqref{eq:Crotor} we can define the bulk, edge, and edge-bulk correlation functions as in 
Sec.~\ref{sec:Cfunctions}.

The boundary criticality of the dissipative O(3) model can be analyzed as before. Results are collected in Fig.~\ref{figSM:rotor} and Table~\ref{tab:scalingdim}. Here we only consider isotropic boundary couplings $\alpha_1 /\alpha = 1$. As for our two previous examples, $\De$
is in good agreement with the prediction of the $\epsilon$ expansion and only experiences weak scaling corrections; we estimate $\De=1.16(2)$ from $C_\mathrm{edge}(\tau=\beta/12)$ and $\De=1.12(2)$ from $C_\mathrm{edge}(\tau=\beta/6)$.
However, as it was already the case for the dimerized chain, $C_\mathrm{edge-bulk}(r)$ experiences a slow crossover with strong scaling corrections.
A power-law fit of the drifting scaling dimension gives $\Db+\De = 1.53(2)$ using a fixed $r=L/24$ and $1.55(3)$ for $r=L/12$. All in all, it is more precise to estimate $\De$ from the edge correlations than from the edge-bulk correlations. It appears to be generic that the latter experience larger scaling corrections. We also note that the error bars obtained from classical Monte Carlo simulations of the dissipative O(3) model are significantly larger than those of our wormhole QMC simulations. Partly, this is because the propagation of the directed loop provides an improved estimator for the spin-spin correlations, 
but it also demonstrates that the wormhole algorithm \cite{PhysRevB.105.165129} is an efficient method to simulate different classes of quantum dissipative systems.

\end{document}